\documentclass[12pt]{article}
\pdfoutput=1

\usepackage{array} 
\usepackage{amssymb}
\usepackage{graphics,graphpap}
\usepackage{graphicx}
\usepackage{color}
\usepackage{graphicx}% Include figure files
\usepackage{dcolumn}% Align table columns on decimal point
\usepackage{epsfig}
\usepackage{epstopdf}
\usepackage{bbm}% bold math
\usepackage{amsmath}
\usepackage{amsfonts}
\usepackage{textcomp}
\usepackage{subcaption}
\usepackage{setspace}
\usepackage{slashed}
\usepackage{multirow}
\usepackage{footnote}
\makesavenoteenv{table}
\usepackage{url}

\newcommand{\beq}{\begin{eqnarray}}
\newcommand{\eeq}{\end{eqnarray}}

\newcommand{\bmp}{\noindent\begin{minipage}{16cm}}
\newcommand{\emp}{\end{minipage}\vskip 7mm} % 7mm untightened

\def\drawbox#1#2{\hrule height#2pt
        \hbox{\vrule width#2pt height#1pt \kern#1pt
              \vrule width#2pt}
              \hrule height#2pt}

\def\Asym#1#2{\vcenter{\vbox{\drawbox{#1}{#2}
              \kern-#2pt % line up boxes
              \drawbox{#1}{#2}}}}

\def\simge{\mathrel{%
   \rlap{\raise 0.511ex \hbox{$>$}}{\lower 0.511ex \hbox{$\sim$}}}}

\def\simle{\mathrel{
   \rlap{\raise 0.511ex \hbox{$<$}}{\lower 0.511ex \hbox{$\sim$}}}}

\def\s#1{\setbox0=\hbox{$#1$}%
\rlap{\ifdim\wd0>.7em\kern.22\wd0\else\kern.1\wd0\fi /}#1}

\usepackage{slashed}

\newcommand{\equref}[1]{Eq.~\eqref{#1}}

\newcommand{\Figref}[1]{Fig.~\ref{#1}}

\newcommand{\tabref}[1]{Tab.~\ref{#1}}
\newcommand{\Tabref}[1]{Tab.~\ref{#1}}

\newcommand{\Secref}[1]{Sec.~\ref{#1}}

\newcommand{\Appref}[1]{Appendix~\ref{#1}}

\newcommand{\sub}[2]{#1_{\mathrm{#2}}} 						%subscript with textmode for the subscript
\newcommand{\ssscript}[3]{#1_{\mathrm{#2}}^{\mathrm{#3}}} 	%sub-super-script with textmode both in the sub ans in the superscript
\newcommand{\ssscriptupper}[3]{#1_{#2}^{\mathrm{#3}}}

\newcommand{\diffd}{\mathrm{d}}													%differential d in non-cursive
\newcommand{\DD}[2]{\frac{\mathrm{d} #1}{\mathrm{d} #2}}						%differential operator with non-cursive d 

\newcommand{\partiald}[1]{\frac{\partial}{\partial #1}}

\newcommand{\Lag}{\mathcal{L}}											%Lagrangian L in special mathcal notation
\newcommand{\LagArg}[1]{\mathcal{L}_{\mathrm{#1}}}											%Lagrangian L of the SM in special mathcal notation
\newcommand{\orderof}[1]{\mathcal{O} \left(#1 \right)}					%order of number in caligraphic or other writing of the ordering symbol
\newcommand{\orderofunit}[2]{\mathcal{O} \left(#1 \, \mathrm{#2} \right)}	%order of number and unit in unit style

\newcommand{\av}[1]{\ensuremath{\left\langle #1 \right\rangle}}
\newcommand{\conj}[1]{\ensuremath{\left(#1\right)^c}}

\newcommand{\hc}{\mathrm{h.c.}} 						%hermitian conjugate in textmode 
\newcommand{\delslashed}{\slashed{\partial}}

\newcommand{\dmf}[1]{dark matter\footnote{}}

\newcommand{\unit}[2]{#1 \, \mathrm{#2}}
\newcommand{\unitonly}[1]{\mathrm{#1}}

\makeatletter
\newcommand*{\textalltt}{}
\DeclareRobustCommand*{\textalltt}{%
	\begingroup
	\let\do\@makeother
	\dospecials
	\catcode`\\=\z@
	\catcode`\{=\@ne
	\catcode`\}=\tw@
	\verbatim@font\@noligs
	\@vobeyspaces
	\frenchspacing
	\@textalltt
}
\newcommand*{\@textalltt}[1]{%
	#1%
	\endgroup
}
\makeatother
\newcommand{\CHP}{\ensuremath{\sub{\mathcal{C}}{HP}}}
\newcommand{\CGamma}{\ensuremath{\sub{\mathcal{C}}{\Gamma}}}

\usepackage{cite}

\begin{document}
%%%%%%%%%%%%%%%%%%%%%%%%%%%%%%%%%%%%%%%%%%%%%%%%%%%%%%%%%%%%%%%%%%%%%%%%%%%

%%%%%%%%%%%%%%%%%%%%%%%%%%%%%%%%%%%%%%%%%%%%%%%%%%%%%%%%%%%%%%%%%%%%%%%%%%%
%%%%%%%%%%%%%%%%%%%%%%%%%%%%%%%%%%%%%%%%%%%%%%%%%%%%%%%%%%%%%%%%%%%%%%%%%%%
\begin{titlepage}
\title{\vspace*{-2.0cm}
\hfill {\small MPP-2015-302}\\[20mm]
\vspace*{-1.5cm}
\bf\Large
Dodelson-Widrow Production of Sterile Neutrino Dark Matter with Non-Trivial Initial Abundance\\[5mm]\ \vspace{-1cm}}

\author{
Alexander Merle$^{a}$\thanks{email: \tt amerle@mpp.mpg.de}~,~~~Aurel Schneider$^{b}$\thanks{email: \tt aurel@physik.uzh.ch}~,~~~and~~Maximilian Totzauer$^{a}$\thanks{email: \tt totzauer@mpp.mpg.de}
\\ \\
$^{a}${\normalsize \it Max-Planck-Institut f\"ur Physik (Werner-Heisenberg-Institut),}\\
{\normalsize \it F\"ohringer Ring 6, 80805 M\"unchen, Germany}\\
$^{b}${\normalsize \it Institute for Computational Science, University of Z\"urich,}\\
{\normalsize \it Winterthurerstrasse 190, CH-8057, Z\"urich Switzerland}\\
}
\date{\today}
\maketitle
\thispagestyle{empty}

\vspace{-0.5cm}
\begin{abstract}
\noindent
The simplest way to create sterile neutrinos in the early Universe is by their admixture to active neutrinos. However, this mechanism, connected to the Dark Matter (DM) problem by Dodelson and Widrow (DW), cannot simultaneously meet the relic abundance constraint as well as bounds from structure formation and X-rays. Nonetheless, unless a symmetry forces active-sterile mixing to vanish exactly, the DW mechanism will unavoidably affect the sterile neutrino DM population created by any other production mechanism. We present a semi-analytic approach to the DW mechanism acting on an arbitrary initial abundance of sterile neutrinos, allowing to combine DW with any other preceding production mechanism in a physical and precise way. While previous analyses usually assumed that the spectra produced by DW and another mechanism can simply be added, we use our semi-analytic results to discuss the validity of this assumption and to quantify its accurateness, thereby also scrutinising the DW spectrum and the derived mass bounds. We then map our results to the case of sterile neutrino DM from the decay of a real SM singlet coupled to the Higgs. Finally, we will investigate aspects of structure formation beyond the usual simple free-streaming estimates in order to judge on the effects of the DW modification on the sterile neutrino DM spectra generated by scalar decay.
\end{abstract}

\end{titlepage}
%%%%%%%%%%%%%%%%%%%%%%%%%%%%%%%%%%%%%%%%%%%%%%%%%%%%%%%%%%%%%%%%%%%%%%%%%%%
%%%%%%%%%%%%%%%%%%%%%%%%%%%%%%%%%%%%%%%%%%%%%%%%%%%%%%%%%%%%%%%%%%%%%%%%%%%

%%%%%%%%%%%%%%%%%%%%%%%%%%%%%%%%%%%%%%%%%%%%%%%%%%%%%%%%%%%%%%%%%%%%%%%%%%%%%%%%%%%%%%%%%%%%%%%%%%
%%%%%%%%%%%%%%%%%%%%%%%%%%%%%%%%%%%%%%%%%%%%%%%%%%%%%%%%%%%%%%%%%%%%%%%%%%%%%%%%%%%%%%%%%%%%%%%%%%
\section{\label{sec:Intro}Introduction}
%%%%%%%%%%%%%%%%%%%%%%%%%%%%%%%%%%%%%%%%%%%%%%%%%%%%%%%%%%%%%%%%%%%%%%%%%%%%%%%%%%%%%%%%%%%%%%%%%%

The biggest mystery of modern cosmology is the identity of Dark Matter (DM). This unknown substance amounts to  $\sim 80\%$ of the matter content in the Universe~\cite{Planck:2015xua}, and it is clear that it was important at the time when galaxies and other large structures have formed. Our best guess for the identity of DM is a new and electrically neutral particle, which is present in sufficient amounts in the Universe. Historically, the most plausible particle physics candidate was a Weakly Interacting Massive Particle (WIMP)~\cite{Lee:1977ua,Bertone:2004pz}. However, up to now we have no unambiguous detection, neither in direct~\cite{Angloher:2014myn,Aprile:2012nq,Akerib:2013tjd} or indirect~\cite{Accardo:2014lma,Ackermann:2013uma,Adriani:2013uda} detection attempts nor by trying to directly produce them at colliders~\cite{ATLAS:2012ky,Chatrchyan:2012me,Khachatryan:2014qwa,Aad:2014vma}. Instead, the limits are pushed towards smaller and smaller couplings.

On the DM side, $N$-body simulations using ordinary cold (i.e., non-relativistic) DM do not perfectly reproduce the Universe at small scales. While there is a rather obvious excess of dwarf galaxy numbers (historically referred to as the \emph{missing satellite} problem~\cite{Klypin:1999uc,Moore:1999nt}, but also present in the field outside of galaxies~\cite{Klypin:2014ira}), more subtle discrepancies related to the internal kinematics of dwarfs, the \emph{cusp-core}~\cite{Dubinski:1991bm,Moore:1994yx} and \emph{too-big-too-fail}~\cite{BoylanKolchin:2011de,Papastergis:2014aba} problems, have been reported as well. Whether these problems can altogether be solved with a proper modeling of gas and stars in cosmological simulations or whether they hint towards a shift of the DM paradigm is still an open question (see, e.g.,~\cite{Weinberg:2013aya} for a review).

A generic non-cold DM candidate is a sterile neutrino with a mass of a few keV -- a heavier and much more feebly interacting version of the active neutrino. It can act as DM if produced in the early Universe in the right amount and with a suitable spectrum. The natural production for this DM is by freeze-in~\cite{McDonald:2001vt,Hall:2009bx} (see also Ref.~\cite{Shakya:2015xnx} for a recent review) driven by the small active components of the predominantly sterile mass eigenstates. It was proposed by Dodelson and Widrow~\cite{Dodelson:1993je} in the context of DM (``DW mechanism''), based on Refs.~\cite{Barbieri:1989ti,Kainulainen:1990ds}. However, given the observational bounds on active-sterile mixing from not observing DM decay\footnote{\emph{Summary of the status of the 3.5~keV line}: Evidence for this X-ray signal has initially been reported by~\cite{Bulbul:2014sua,Boyarsky:2014jta} using XMM-Newton and Chandra data for several clusters as well as for Andromeda. Three main criticisms have been raised: 1.) The signal does not show up in all data sets~\cite{Riemer-Sorensen:2014yda,Anderson:2014tza,Malyshev:2014xqa,Urban:2014yda,Tamura:2014mta,Jeltema:2015mee} -- although some groups found new versions of the line~\cite{Urban:2014yda,Iakubovskyi:2015wma}. 2.) The emission lines of chemical elements are not treated correctly~\cite{Jeltema:2014qfa,Jeltema:2014mla,Phillips:2015wla}. 3.) The signal does not follow the generic DM profiles at the centres of galaxies~\cite{Urban:2014yda,Carlson:2014lla}. All these arguments have also been criticised~\cite{Anderson:2014tza,Boyarsky:2014ska,Boyarsky:2014paa,Bulbul:2014ala}, and at the moment it is not clear what the final outcome will be. On top, also the original authors scrutinise their observations~\cite{Iakubovskyi:2015dna,Iakubovskyi:2015kwa}, which will hopefully clarify some aspects. In general, however, we should keep in mind that the ``signal'' is suffering from low statistics, which implies that \emph{both} its ``detection''  and ``exclusion'' depend on the data sets used and analysis applied. Finally, future observations from Astro-H~\cite{Koyama:2014zca,Kitayama:2014fda}, LOFT~\cite{LOFT,Zane:2014vya}, eROSITA~\cite{Zandanel:2015xca}, or searches using X-ray microcalorimeter sounding rockets~\cite{Figueroa-Feliciano:2015gwa} could soon clear up the situation~\cite{Iakubovskyi:2015kwa}.} as well as the relatively hot spectrum resulting from the DW mechanism, this simple way of producing sterile neutrino DM is excluded~\cite{Seljak:2006qw,Viel:2013apy}.

With the most generic possibility excluded, alternatives are pursued. Resonant production (``Shi-Fuller (SF) mechanism''~\cite{Shi:1998km}, see~\cite{Enqvist:1990ek} for an earlier work) has been investigated heavily~\cite{Abazajian:2001nj,Asaka:2006nq,Laine:2008pg,Canetti:2012kh,Abazajian:2014gza,Venumadhav:2015pla}. It relies on a primordial lepton number asymmetry and leads to a colder spectrum consistent with structure formation. Another way is to produce keV neutrinos by decays of other particles. These can be inflatons~\cite{Shaposhnikov:2006xi,Bezrukov:2009yw,Bezrukov:2014nza}, electrically neutral~\cite{Kusenko:2006rh,Petraki:2007gq,Merle:2013wta,Kang:2014cia,Merle:2015oja,Humbert:2015epa} or charged~\cite{Frigerio:2014ifa} scalars, other spin states like light vectors~\cite{Boyanovsky:2008nc,Shuve:2014doa} or Dirac fermions~\cite{Abada:2014zra}, or known particles like the pion~\cite{Lello:2014yha,Lello:2015uma}.\footnote{For parent particles with gauge charges, thermal corrections may be non-negligible~\cite{Drewes:2015eoa}.} While both, SF and decay production, are consistent at this stage, they could be distinguished with improved data on structure formation~\cite{Merle:2014xpa}. There exists a further mechanism: if sterile neutrinos are charged under a new interaction beyond the Standard Model (SM), they may after all be produced by freeze-out~\cite{Bezrukov:2009th,Nemevsek:2012cd} -- however, the resulting overabundance has to be corrected by producing additional entropy~\cite{Scherrer:1984fd,Scherrer:1987rr}, creating tension with big bang nucleosynthesis~\cite{King:2012wg}.

In this paper, we will re-investigate the DW mechanism. There are several reasons to do so. First, one often encounters incorrect statements about DW, e.g., that it would produce a spectrum of thermal shape~\cite{Dodelson:1993je,Colombi:1995ze}. This assumption is even used when deriving consequences for structure formation~\cite{Viel:2005qj,Herpich:2013yga,Menci:2012kk,Lovell:2013ola}, and it also enters the exclusion itself~\cite{Viel:2013apy,Baur:2015jsy}. Second, while not all DM can be produced by DW, the mechanism cannot be switched off as long as there is mixing between active and sterile neutrinos (which is generic, see e.g.\ Refs.~\cite{Shaposhnikov:2006nn,Cogollo:2009yi,Kusenko:2010ik,Lindner:2010wr,Merle:2011yv,Adulpravitchai:2011rq,Zhang:2011vh,Araki:2011zg}, or Ref.~\cite{Merle:2013gea} for a review). Thus, virtually \emph{any} production mechanism will experience a modulation by the DW mechanism. This is the case for many settings investigated recently~\cite{Abada:2014zra,Merle:2014xpa,Kang:2014cia,Humbert:2015epa}, yet the effect has not been taken into account. Third, we present the explicit example of scalar decay production being affected by the DW contribution. Decay production has been studied earlier by two of us (AM \&\ MT)~\cite{Merle:2015oja}, so that the consequences of the DW modification do not only serve as an illustration for the current paper, but also as a-posteriori justification of our previous results.

This paper is structured as follows. In \Secref{sec:DWInitialAbundance} we present a formal but general solution for the DM-distribution produced from or modulated by the DW mechanism, valid for \emph{any} initial spectrum. We then use this formal solution in \Secref{sec:NumericalAnalysisDW} to derive an absolute upper bound on the strength of the DW-modification, which is strongly constrained by not observing X-rays from sterile neutrino decay. In \Secref{sec:StructureFormation} we discuss quite generally the implications of the number of satellite galaxies on the maximal allowed fraction that DW produced sterile neutrinos can contribute to DM, irrespective of what makes up the rest. As a concrete example for a modulation of the DM distribution function by DW, we use \Secref{sec:SDCase} to compute the effect on singlet scalar decay production. We will also exemplify some structure formation aspects.\footnote{This justifies neglecting the DW contribution, as done earlier by two of us (AM \&\ MT) in Ref.~\cite{Merle:2015oja}.} We conclude in \Secref{sec:Conclusion}. Technical details, such as a formal proof of the DW-consistency relation can be found in Appendix~\ref{app:A:FormalSolution}.

%%%%%%%%%%%%%%%%%%%%%%%%%%%%%%%%%%%%%%%%%%%%%%%%%%%%%%%%%%%%%%%%%%%%%%%%%%%%%%%%%%%%%%%%%%%%%%%%%%
%%%%%%%%%%%%%%%%%%%%%%%%%%%%%%%%%%%%%%%%%%%%%%%%%%%%%%%%%%%%%%%%%%%%%%%%%%%%%%%%%%%%%%%%%%%%%%%%%%
\section{\label{sec:DWInitialAbundance}DW mechanism with non-zero initial abundance}
%%%%%%%%%%%%%%%%%%%%%%%%%%%%%%%%%%%%%%%%%%%%%%%%%%%%%%%%%%%%%%%%%%%%%%%%%%%%%%%%%%%%%%%%%%%%%%%%%%

In this section, we present the most general solution to the Boltzmann equation describing the DW mechanism. Although the solution is formal, as it contains integrals which are not yet evaluated as they depend on the exact parameters and functions inserted, our solution allows to get an intuitive picture of the workings behind the mechanism.

In the generic setup used for DW production, the SM is extended by $n_R$ right-handed (RH) fermion singlets $N_i$, where $i=1,...,n_R$. In fact, given that at least two light neutrinos are massive~\cite{Bergstrom:2015rba,Gonzalez-Garcia:2014bfa}, we require at least two such RH neutrinos, $n\geq 2$~\cite{Schechter:1980gr,Xing:2007uq}, or possibly even more such as the three fields considered in the neutrino-minimal SM ($\nu$MSM)~\cite{Asaka:2005an}. However, when aiming at DW production only, we can work with $n_R=1$ for simplicity. This will not affect the results but allow us to concentrate on the actual mechanism. The inclined reader can find a more general discussion in Appendix~\ref{app:A:FormalSolution:DWChoiceNR}.

With $n_R=1$, the following new terms can appear in the Lagrangian:
\begin{align}
  \Lag \supset -L_\alpha \tilde{H} y_{\alpha} N_{1} - \frac{1}{2} \overline{\conj{N_1}} M_{1} N_{1} + \hc 
  \label{eq:LagrangianTerms}
\end{align}
We have introduced four new parameters: the Majorana mass $M_1$ and three Yukawa couplings $y_\alpha$ with $\alpha \in \{e, \mu,\tau\}$. Instead of the Yukawas, however, we will use the \emph{active-sterile mixing angles} $\left(\theta^e,\theta^\mu,\theta^\tau\right)$:
\begin{align}
  \theta^\alpha \equiv \frac{y_\alpha \sub{v}{EW}}{M_1} \,,
  \label{eq:DefineMixingAnglesNR1}
\end{align}
where $\sub{v}{EW}$ is the vacuum expectation value of the SM Higgs field. Note that \equref{eq:DefineMixingAnglesNR1} is in principle only valid for small mixing, i.e.~$y_\alpha \sub{v}{EW}/M_1 \ll 1$. However, in practice this condition is always fulfilled due to the X-ray bound~\cite{Merle:2012xq}. The three mixing angles $\left(\theta^e,\theta^\mu,\theta^\tau\right)$, although small, are the driving forces behind $N_1$-production: their values ultimately decide about how likely a certain process involving SM particles may produce a sterile neutrino.

%%%%%%%%%%%%%%%%%%%%%%%%%%%%%%%%%%%%%%%%%%%%%%%%%%%%%%%%%%%%%%%%%%%%%%%%%%%%%%%%%%%%%%%%%%%%%%%%%%
\subsection{\label{sec:DWInitialAbundance:FormalSolution}The full DW Boltzmann equation and its formal solution}
%%%%%%%%%%%%%%%%%%%%%%%%%%%%%%%%%%%%%%%%%%%%%%%%%%%%%%%%%%%%%%%%%%%%%%%%%%%%%%%%%%%%%%%%%%%%%%%%%%

The fundamental quantity describing the properties of a particle species in the Universe is its momentum distribution function $f(p,t)$, abbreviated as MDF. It is typically governed by a set of Boltzmann equations. In the epoch we are interested in, well before the decoupling of the cosmic microwave background, the Universe is homogeneous and isotropic. Therefore, the MDF will only depend on the modulus $p$ of the momentum and on time $t$ or, equivalently, on the temperature $T$: $f=f(p,T)$. Note, however, that the temperature $T$ denotes the temperature of the photons that stay in equilibrium until CMB decoupling. In that sense, $T$ should indeed be interpreted as a global cosmic evolution variable with a one-to-one correspondence to cosmic time $t$ and \emph{not necessarily} as a thermodynamic property of the species described by the distribution function $f$. 

In a Friedman-Robertson-Walker Universe, the Boltzmann equation describing the dynamics of the sterile neutrino -- and thus yielding the distribution $f_N (p,T)$ -- reads:
\begin{align}
 \left(\DD{T}{t} \partiald{T} - H p \partiald{p}\right) f_N\left(T,p\right) = \sum\limits_{i}{C_i\left[f_{\beta_i}\right]}\;.
 \label{eq:BoltzmannFRWTempMomentum}
\end{align}
Here, $H=H\left(T\right)$ is the Hubble function and the \emph{collision terms} ${C_i\left[f_{\beta_i}\right]}$ encode all production and/or annihilation channels\footnote{In principle, also scattering channels have to be accounted for in the collision terms. While not changing the number density of a species, scatterings may change the distribution of the momentum modulus $p$. In cases where \equref{eq:BoltzmannFRWTempMomentum} will finally be integrated to obtain a Boltzmann equation on the level of particle number densities, scatterings are usually neglected, though they can have an effect in theory.} (indexed by $i$) of the species of interest. Each collision term is a functional of the distribution functions of all species $\beta_i$ taking part in process $i$. For example, if a sterile neutrino is produced by a $W$-boson decaying into a charged lepton, $W^- \to l^- N_1$, then the distribution functions of both $W^-$ and $l^-$ enter the equations. While in general one would have to compute all of them, in the early Universe it is often clear that most species follow their analytically known equilibrium distributions. This way, one usually only needs a small number of coupled equations of the type of \equref{eq:BoltzmannFRWTempMomentum}. Still, one should be very careful with this choice if it is a priori not clear in which range of the cosmic evolution the system of coupled equations varies significantly.

Dividing \equref{eq:BoltzmannFRWTempMomentum} by the temperature-time derivative and inserting the DW collision term in an abstract form, we arrive at:
\begin{align}
  \left(\partiald{T} - \kappa\left(T\right)\ p \partiald{p}\right)f_N \left(T,p\right) = h\left(T,p\right) \left[ \sub{f}{th}\left(T,p\right) - f_N\left(T,p\right)\right]\;,
 \label{eq:BoltzmannDWAbstract}
\end{align}
where have defined the \emph{redshift integrand} $\kappa\left(T\right)$ to be:
\begin{align}
 \kappa\left(T\right) = H\left(T\right) \DD{t}{T} \;.
 \label{eq:Def:kappa}
\end{align}
In Eq.~\eqref{eq:BoltzmannDWAbstract}, $\sub{f}{th}$ denotes a thermal Fermi-Dirac distribution and $h$, explicitly displayed in Eq.~\eqref{eq:h-explicit}, encodes all details of the conversion of active to sterile neutrinos. This latter quantity is where all the physics enters, and it will be discussed in detail in Appendix~\ref{app:A:FormalSolution}. It is also discussed at length in the literature, see e.g.~\cite{Abazajian:2001nj,Venumadhav:2015pla}. Note that $h=h[T,p, M_1, (\theta^e,\theta^\mu,\theta^\tau)]$ also depends on the mass $M_1$ of the sterile neutrino and on all three mixing angles $(\theta^e,\theta^\mu,\theta^\tau)$. For the sake of clarity, we will however suppress these arguments whenever there is no risk of confusion.

The compact notation allows to grasp the essential parts of the Boltzmann equation for the sterile neutrino. Since the active-to-sterile conversion is a $1 \leftrightarrow 1$ process, it is clear that the Boltzmann equation must depend linearly on both the MDFs of the sterile and active neutrinos (the latter being given by $\sub{f}{th}$ as long as active neutrinos are in thermal equilibrium). The right-hand side of \equref{eq:BoltzmannDWAbstract} can be interpreted as the sum of a
\begin{eqnarray}
 &&\text{\emph{gain term}:}\ \ \ h\left(T,p\right) \sub{f}{th}\left(T,p\right), \ \ \ \text{and a}\nonumber\\
 &&\text{\emph{loss term}:}\ \ \  -h\left(T,p\right) f_{N}\left(T,p\right)\,.
\end{eqnarray}

In the standard DW scenario \emph{without} any initial abundance, the loss term is usually neglected,\footnote{It can, however, not be neglected in the case of \emph{resonant} active-sterile conversion, even for vanishing initial abundance, see Ref.~\cite{Abazajian:2001nj}.} which greatly simplifies the computation. We will keep the term, though, thus allowing for an arbitrary distribution function of sterile neutrinos produced from another production mechanism before the onset of DW, which becomes efficient only at temperatures of $\orderofunit{100}{MeV}$~\cite{Abazajian:2001nj}.

Before advancing to the full solution of Eq.~\eqref{eq:BoltzmannDWAbstract}, let us investigate the properties of $\kappa\left(T\right)$. Conservation of the comoving entropy density can be cast as $g_S\left(T\right) T^3 a^3\left(T\right) = \mathrm{const.}$, where $g_S\left(T\right)$ is the number of effective entropy degrees of freedom (d.o.f.). Differentiating this equation with respect to time and changing variables from $t$ to $T$ yields:
\begin{align}\
 \kappa\left(T\right) = -\frac{1}{T}\left(1 + \frac{1}{3} \frac{T g_S'}{g_S}\right) \: ,
 \label{eq:kappa:EntropyVersion}
\end{align}
where $g_S'$ denotes the derivative of $g_S$ with respect to $T$. By virtue of \equref{eq:kappa:EntropyVersion}, we can immediately derive:
\begin{align}
 \exp\left(\int\limits_{T_a}^{T_b}{\diffd T_1 \kappa\left(T_1\right)}\right) = \frac{T_a}{T_b} \left(\frac{g_S\left(T_a\right)}{g_S\left(T_b\right)}\right)^{1/3} \;,
 \label{eq:RedshiftIntegral}
\end{align}
which will be used to compactify our final solution. The above result justifies the name ``redshift integrand'' for $\kappa$: in a \emph{collisionless} Boltzmann equation, the solution of \equref{eq:BoltzmannFRWTempMomentum} has to account for redshift only, and for an initial distribution $\sub{f}{ini}(p)$ it reads\footnote{Here, we suppress the time-like argument $T$ in the definition of $\sub{f}{ini}$. This quantity is given as an input to the equations at a given initial temperature $T=\sub{T}{ini}$, such that all further dependence on the temperature $T$ of the Universe is described exclusively by the evolution equations.}

\begin{align}
 \sub{f}{collisionless}\left(p,T\right)=\sub{f}{ini}\left(\frac{a\left(T\right)}{a\left(\sub{T}{ini}\right)} p \right) = \sub{f}{ini}\left(\frac{\sub{T}{ini}}{T}\left(\frac{g_S\left(\sub{T}{ini}\right)}{g_S\left(T\right)}\right)^{1/3}p\right)\;,
 \label{eq:SolutionCollisionlessBoltzmannRedshift}
\end{align}
indeed fulfilling the boundary condition $\sub{f}{collisionless}\left(p,\sub{T}{ini}\right)=\sub{f}{ini}\left(p\right)$. The exponentiated integral of $\kappa$ turns out to be precisely the factor that accounts for the redshift of a collisionless species. If the number of d.o.f.~stays constant, the expected approximate redshift proportionality to $T^{-1}$ is recovered, too.

Let us now proceed to the solution of \equref{eq:BoltzmannDWAbstract}. In its most condensed form, the solution at some final temperature $\sub{T}{f}$ reads
\begin{equation}
 f_{N}\left(\sub{T}{f},p\right) = \mathcal{S}\left(\sub{T}{f},\sub{T}{ini},\sub{T}{f},p,\right) \; \left[\sub{f}{ini}\left(\frac{\sub{T}{ini}}{\sub{T}{f}} \left(\frac{g_S\left(\sub{T}{ini}\right)}{g_S\left(\sub{T}{f}\right)}\right)^{1/3}p\right) + \sub{f}{DW}\left(\sub{T}{f},\sub{T}{ini},p\right) \right]\;,
 \label{eq:SolutionExactCondensed}
\end{equation}
where we have made use of the following abbreviations:
\begin{align}
 \mathcal{S}\left(T_a,T_b,T_c,p\right)&\equiv\exp\left[\int\limits_{T_a}^{T_b}{\diffd T_2\ h\left(T_2, \frac{T_2}{T_c}\left(\frac{g_S\left(T_2\right)}{g_S\left(T_c\right)}\right)^{1/3}p\right)}\right] \; ,
 \label{eq:Def:S} 
 \\
 \sub{f}{DW}\left(T_a,T_b,p\right) &\equiv - \int\limits_{T_a}^{T_b}{\diffd T_2 \mathcal{S}^{-1}\left(T_2,T_b,T_a,p\right)
 \left(h\sub{f}{th}\right) \left(T_2,\frac{T_2}{\sub{T}{f}} \left(\frac{T_2}{\sub{T}{f}} \frac{g_S\left(T_2\right)}{g_S\left(\sub{T}{f}\right)}\right)^{1/3}p\right)} \;.
 %\sub{f}{th}\left(T_2,\frac{T_2}{\sub{T}{f}} \left( \frac{g_S\left(T_2\right)}{g_S\left(\sub{T}{f}\right)}\right)^{1/3}p\right) h\left(T_2,\frac{T_2}{\sub{T}{f}} \left( \frac{g_S\left(T_2\right)}{g_S\left(\sub{T}{f}\right)}\right)^{1/3}p\right)}\;,
 \label{eq:Def:fDW}
\end{align}
In \equref{eq:Def:fDW}, we have introduced the notation $\left(h \sub{f}{th}\right)\left(T, p\right) \equiv h\left(T, p\right) \sub{f}{th}\left(T, p\right)$. As before, we have suppressed the arguments $M_1$ and $\left(\theta^e,\theta^\mu,\theta^\tau\right)$. Note that \equref{eq:SolutionExactCondensed} trivially fulfills the boundary condition $f_N\left(\sub{T}{ini},p\right) = \sub{f}{ini}\left(p\right)$, as it should. The prefactor $\mathcal{S}$ can be interpreted as \emph{damping factor} converting part of the initial distribution $\sub{f}{ini}$ into active neutrinos by active-sterile conversion \emph{in addition} to redshifting the distribution. The product of $\mathcal{S}\sub{f}{DW}$ can be interpreted as the pure DW contribution, resulting from a vanishing initial abundance.

\paragraph{Consistency check} The form of \equref{eq:SolutionExactCondensed} is most suitable to discuss an important consistency check of the solution: \emph{the MDF at some temperature $T_1$ has no memory of the dynamics that shaped it at temperatures $T>T_1$.} Accordingly, the temperature $\sub{T}{ini}$ can be chosen \emph{arbitrarily} if the DW contribution produced at $T>\sub{T}{ini}$ is included into $\sub{f}{ini}$. To be more concrete, we expect the following relation to hold:
\begin{eqnarray}
  && \mathcal{S}\left(\sub{T}{f},\sub{T}{ini},\sub{T}{f},p\right) \sub{f}{DW}\left(\sub{T}{f},\sub{T}{ini},p\right) \stackrel{!}{=} \nonumber\\
  && \mathcal{S}\left(\sub{T}{f},T_3,\sub{T}{f},p\right) \Bigg[\mathcal{S}\left(\sub{T}{3},\sub{T}{ini},T_3,\frac{T_3}{\sub{T}{f}} \left(\frac{g_S\left(T_3\right)}{g_S\left(\sub{T}{f}\right)}\right)^{1/3}p\right) \sub{f}{DW}\left(T_3,\sub{T}{ini},\frac{T_3}{\sub{T}{f}} \left(\frac{g_S\left(T_3\right)}{g_S\left(\sub{T}{f}\right)}\right)^{1/3}p\right) + \nonumber\\ 
  && \quad\quad\quad\quad\quad\quad\quad\quad \sub{f}{DW}\left(\sub{T}{f},T_3,p\right)\Bigg]\;, 
 \label{eq:SolutionConservationRule}
\end{eqnarray}
for arbitrary $T_3 \in \left[T_f,\sub{T}{ini}\right]$. This equation states that sterile neutrinos produced via DW until some temperature $T_3$ can just be re-interpreted as \emph{initial} abundance $\sub{f}{ini}$ present at $T_3$; in fact, they could have been produced by \emph{any} mechanism. But, ``stopping the clock'' at $T_3$ and re-starting it, they can be used as a starting condition for DW production from temperature $T_3$ onwards. This \emph{oblivion} of the initial MDF on how it arose in the first place is an important feature of the solution described by Eqs.~\eqref{eq:SolutionExactCondensed}--\eqref{eq:Def:fDW}, and it can be used to double-check it. While the physics of this relation should be intuitively clear, we will present a formal-analytic proof in \Appref{app:A:FormalSolution:ConsitencyRelationProof}.

%%%%%%%%%%%%%%%%%%%%%%%%%%%%%%%%%%%%%%%%%%%%%%%%%%%%%%%%%%%%%%%%%%%%%%%%%%%%%%%%%%%%%%%%%%%%%%%%%%
\subsection{\label{sec:DWInitialAbundance:StandardDW}A note on the DW case without initial abundance}
%%%%%%%%%%%%%%%%%%%%%%%%%%%%%%%%%%%%%%%%%%%%%%%%%%%%%%%%%%%%%%%%%%%%%%%%%%%%%%%%%%%%%%%%%%%%%%%%%%

Though being ruled out as sole production mechanism~\cite{Viel:2013apy}, we have argued before that DW can contribute subdominantly to the relic abundance, e.g.\ in a mixed DM setting. Some features attributed to pure DW production in the literature are in fact not correct. In particular the close-to-thermal shape noted in~\cite{Dodelson:1993je,Colombi:1995ze} has been used to exclude the DW mechanism in the first place. But, while the DW mechanism is in any case excluded as only source of sterile neutrino DM, the current bounds can actually be enhanced if one takes into account the correct spectral shape.

It is worthwhile to use our semi-analytical results to discuss the case of DW production without initial abundance in a precise and physical way, in order to probe the quality of the approximations used earlier.  We will in particular discuss why a suppressed thermal shape of the DW spectrum, as often adopted in the literature, is in fact \emph{not} a very accurate estimate, especially if the high momentum part of the distribution is important -- like in analyses concerning cosmological structure formation where precisely that part puts the DW mechanism into trouble. 

\begin{figure}
 %\centering
 \hspace{-0.5cm}
 \begin{tabular}{lr}
 \includegraphics[width=8.3cm]{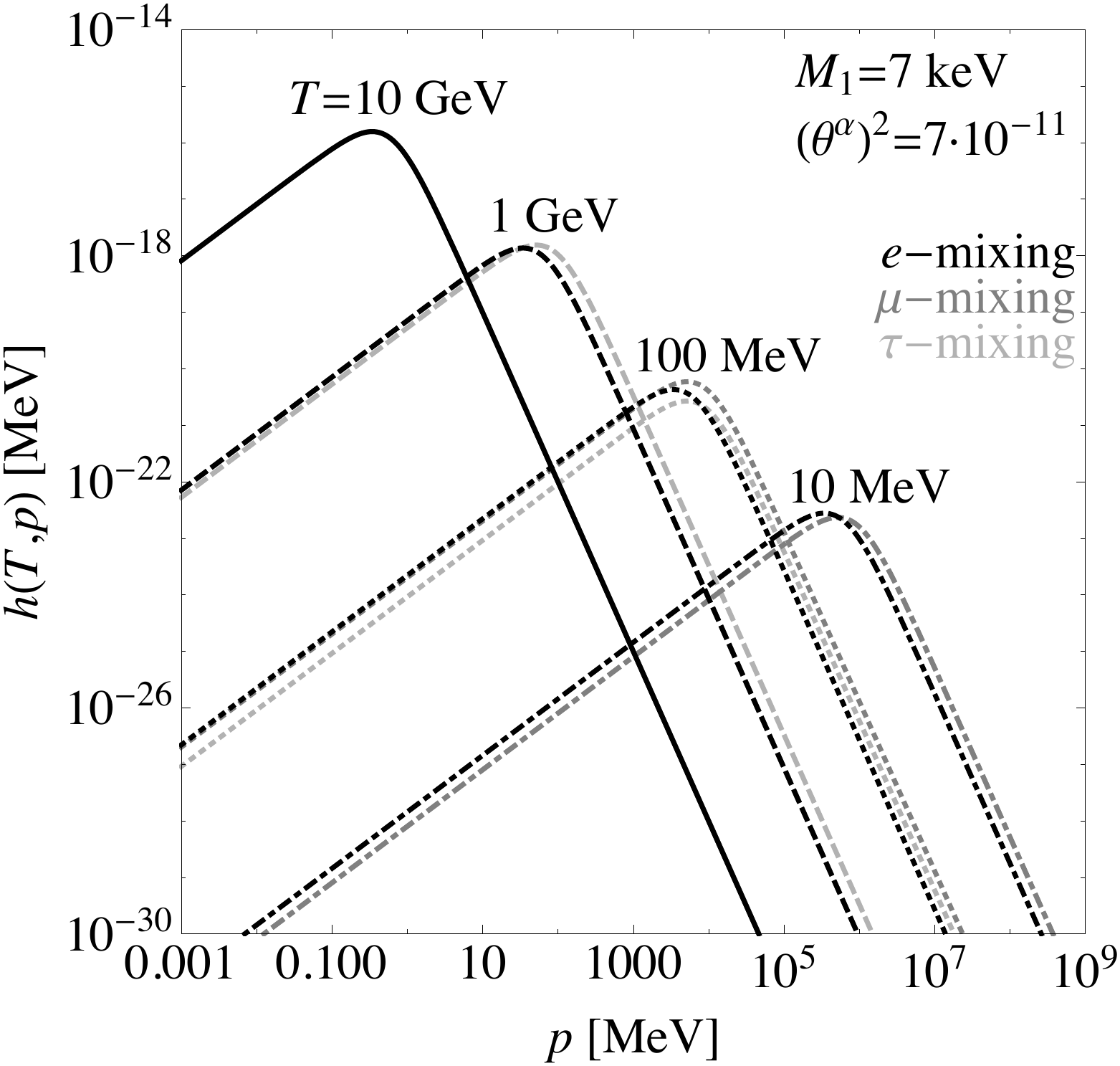} & \includegraphics[width=8cm]{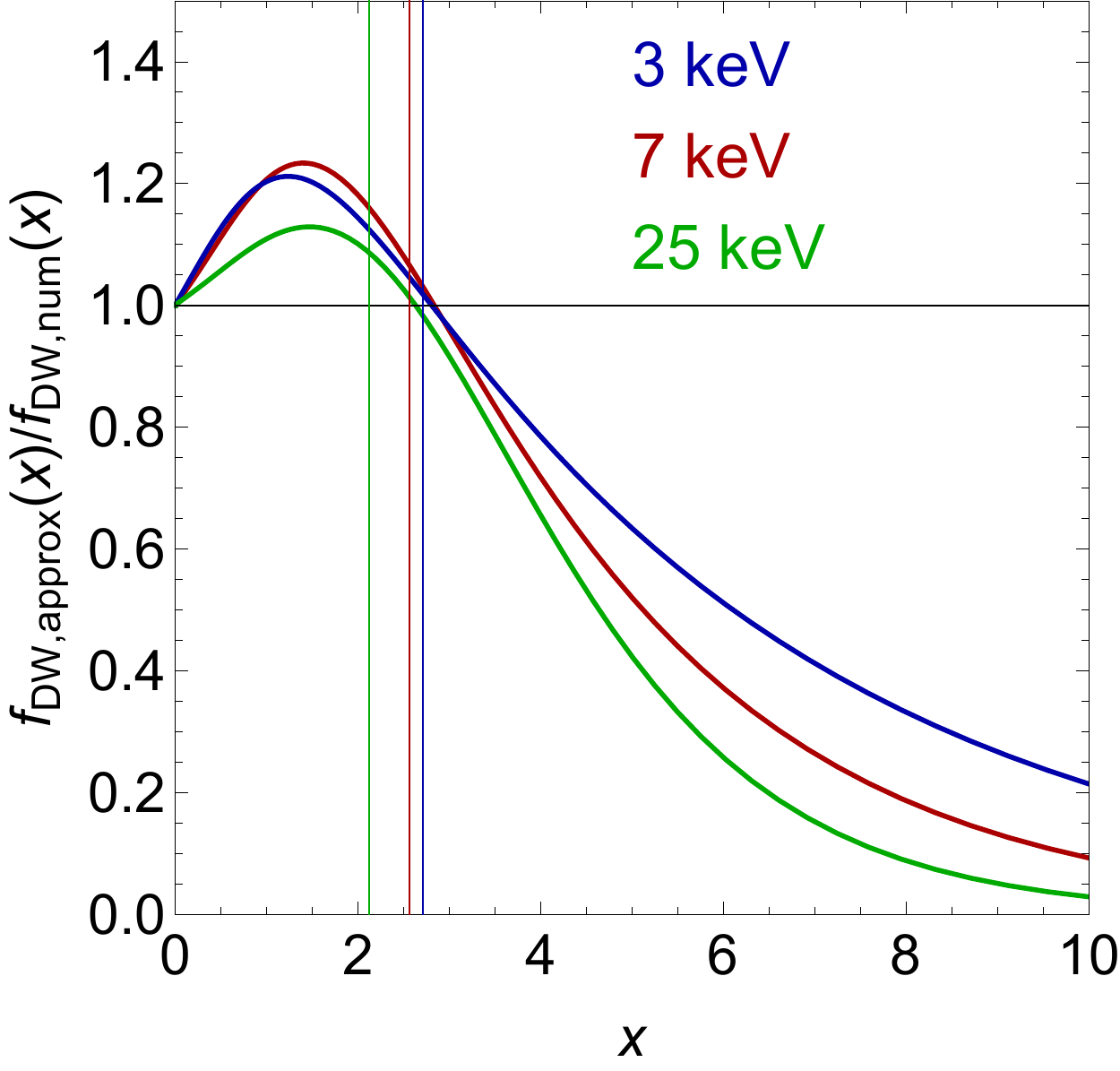}
 \end{tabular}
 \caption{\label{fig:h-variance} \emph{Left panel}: The function $h$ changes dramatically with $p$. \emph{Right panel}: Deviation of the best \emph{ex-post} chosen approximation from the numerical result in the case of pure $e$-mixing. The vertical lines show the average momentum of the numerical distribution to give an indication, where the deviation is most relevant.}
\end{figure}

To do so, let us solve \equref{eq:BoltzmannDWAbstract} neglecting the term $-h\left(T,p\right) f_N\left(T,p\right)$ on the right-hand side. Then, the solution at temperature $\sub{T}{f}$, as derived from Eq.~\eqref{eq:SolutionExactCondensed}, reads:
\begin{align}
 f_N^{\mathrm{DW}}\left(T,p\right) =  \int\limits_{\sub{T}{ini}}^{\sub{T}{f}}{\diffd T_2\ \sub{f}{th}\left(T_2, \frac{T_2}{\sub{T}{f}} \left(\frac{g_S\left(T_2\right)}{g_S\left(\sub{T}{f}\right)}\right)^{1/3}p\right)h\left(T_2, \frac{T_2}{\sub{T}{f}} \left(\frac{g_S\left(T_2\right)}{g_S\left(\sub{T}{f}\right)}\right)^{1/3}p\right)}\;.
 \label{eq:SolutionLowInitial}
\end{align}
Since a thermal distribution of a nearly massless species just depends on the ratio $p/T$ of momentum and temperature, $\sub{f}{th}$ in \equref{eq:SolutionLowInitial} depends on $T_2$ only via the term $g_S\left(T_2\right)$. Thus, \emph{only if $g_S$ varied sufficiently slowly with $T_2$}, one could replace $g_S\left(T_2\right)$ by some average value $\av{g_S}$ and pull the thermal part $\sub{f}{th}$ in front of the integral, hence resembling the shape of a thermal distribution (only multiplied by a suppression factor). \emph{If} one can do that, the solution to \equref{eq:BoltzmannDWAbstract} is indeed given by:
\begin{align}
 f_N^\mathrm{DW}\left(\sub{T}{f},p\right) \approx  \frac{1}{\exp \left( \frac{p}{\sub{T}{f}} \left(\frac{\av{g_S}}{g_S\left(\sub{T}{f}\right)}\right)^{1/3} \right) + 1} \;\; \int\limits_{\sub{T}{ini}}^{\sub{T}{f}}{\diffd T_2\ h\left(T_2, \frac{T_2}{\sub{T}{f}} \left(\frac{g_S\left(T_2\right)}{g_S\left(\sub{T}{f}\right)}\right)^{1/3}p\right)}\;.
 \label{eq:ThermalApprox}
\end{align}
Even if that was the case, which is however certainly not true at least during the QCD transition which generically happens around the peak of DW production~\cite{Barbieri:1989ti,Dodelson:1993je}, for the resulting distribution to be of thermal shape the function $h$ would \emph{in addition} need to vary only very slowly with the momentum $p$. But, plotting $h$ as a function of $p$ for different values of $T_2$, cf.\ left panel of \Figref{fig:h-variance}, we see that this is not at all the case. Accordingly, the statement about the distribution being of thermal shape (only redshifted and with a suppression factor) is just \emph{not correct}. 

To quantify the discrepancy between the approximation and the numerical result, we have chosen three benchmark cases: $M_1=\unit{3}{keV}$, $M_1=\unit{7}{keV}$, and $M_1=\unit{25}{keV}$. We have then computed the ratios between the approximate result in \equref{eq:ThermalApprox} and the exact one in \equref{eq:SolutionLowInitial}  for pure $e$-mixing (the results for pure $\mu$- or $\tau$-mixing are very similar). The freedom of choosing $\av{g_S}$ was eliminated by fixing this parameter such that the particle number density is equal to the numerical result, which -- of course -- is unknown in case one uses the approximation only! In this sense, our \emph{ex-post} choice of $\av{g_S}$ yields the best approximation possible, and even this is not very accurate for high momenta. The right panel of \Figref{fig:h-variance} illustrates the deviation for the three benchmark cases as a function of rescaled momentum $x = p/T$. Note that these benchmark cases are valid for \emph{any} choice of the mixing angle, since by virtue of Eqs.~\eqref{eq:SolutionLowInitial} and~\eqref{eq:ThermalApprox} both the approximation and the numerical result with vanishing initial abundance are exactly proportional to $\sin^2\left(2\theta\right)$.

The right panel of \Figref{fig:h-variance} reveals that the approximation is perfect for $x \rightarrow 0$, which is due to the fact that the Fermi-Dirac distribution tends to $1/2$, irrespective of the choice of $\av{g_S}$. However, the thermal shape approximation systematically underestimates the \emph{high} momentum modes, which are in fact the most decisive ones when excluding the DW mechanism by cosmic structure formation. 

\begin{figure}
 \centering
 \begin{subfigure}{0.32 \textwidth}
 \includegraphics[width=\textwidth]{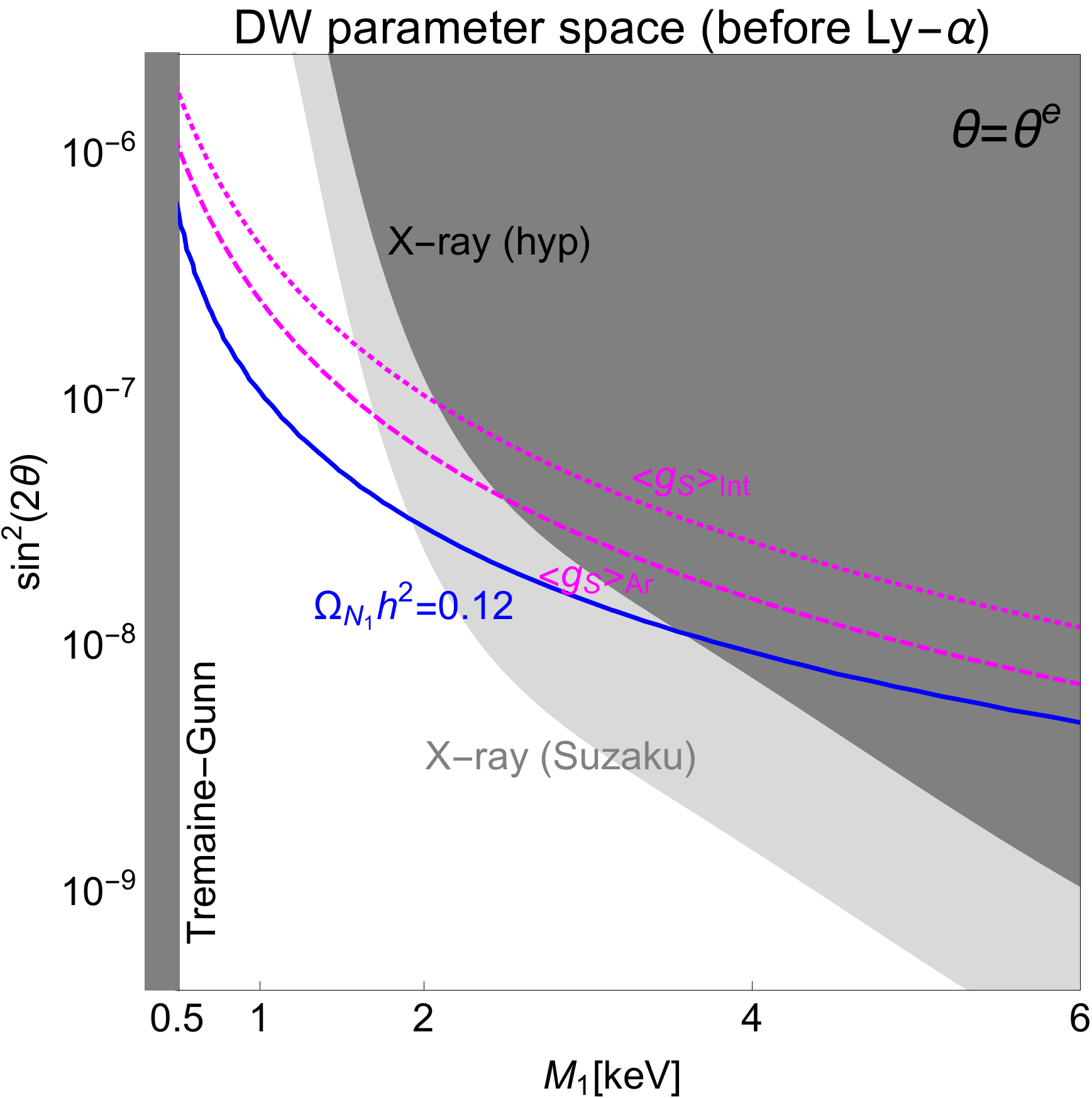}
 \caption{$\theta^e=\ssscript{\theta}{max}{hyp}\left(M_1\right)$.}
 \label{fig:ThetaMPlanePureDWIsoabundance:e}
 \end{subfigure}
 \begin{subfigure}{0.32 \textwidth}
 \includegraphics[width=\textwidth]{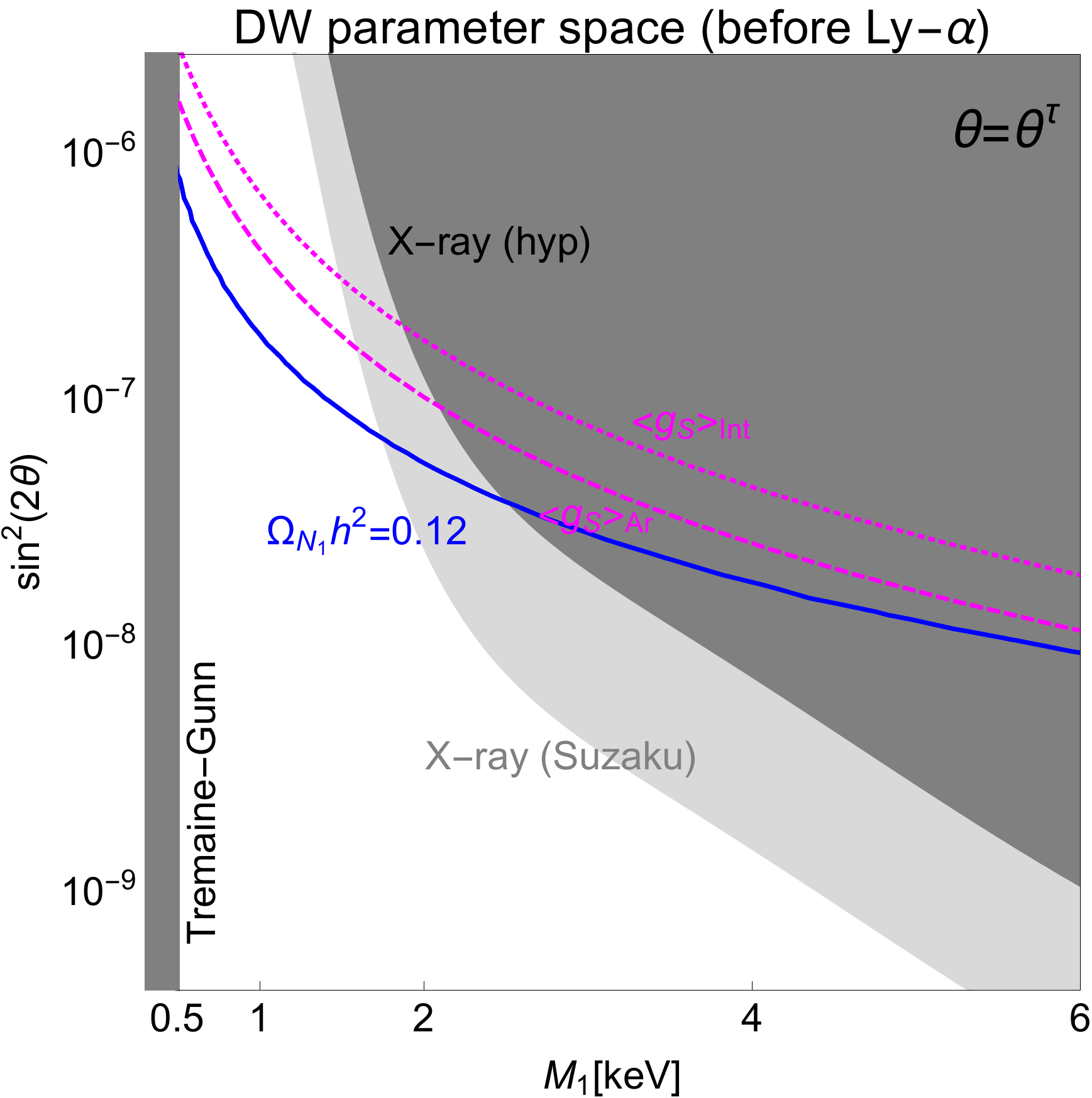}
 \caption{$\theta^\mu=\ssscript{\theta}{max}{hyp}\left(M_1\right)$.}
 \end{subfigure}
 \begin{subfigure}{0.32 \textwidth}
 \includegraphics[width=\textwidth]{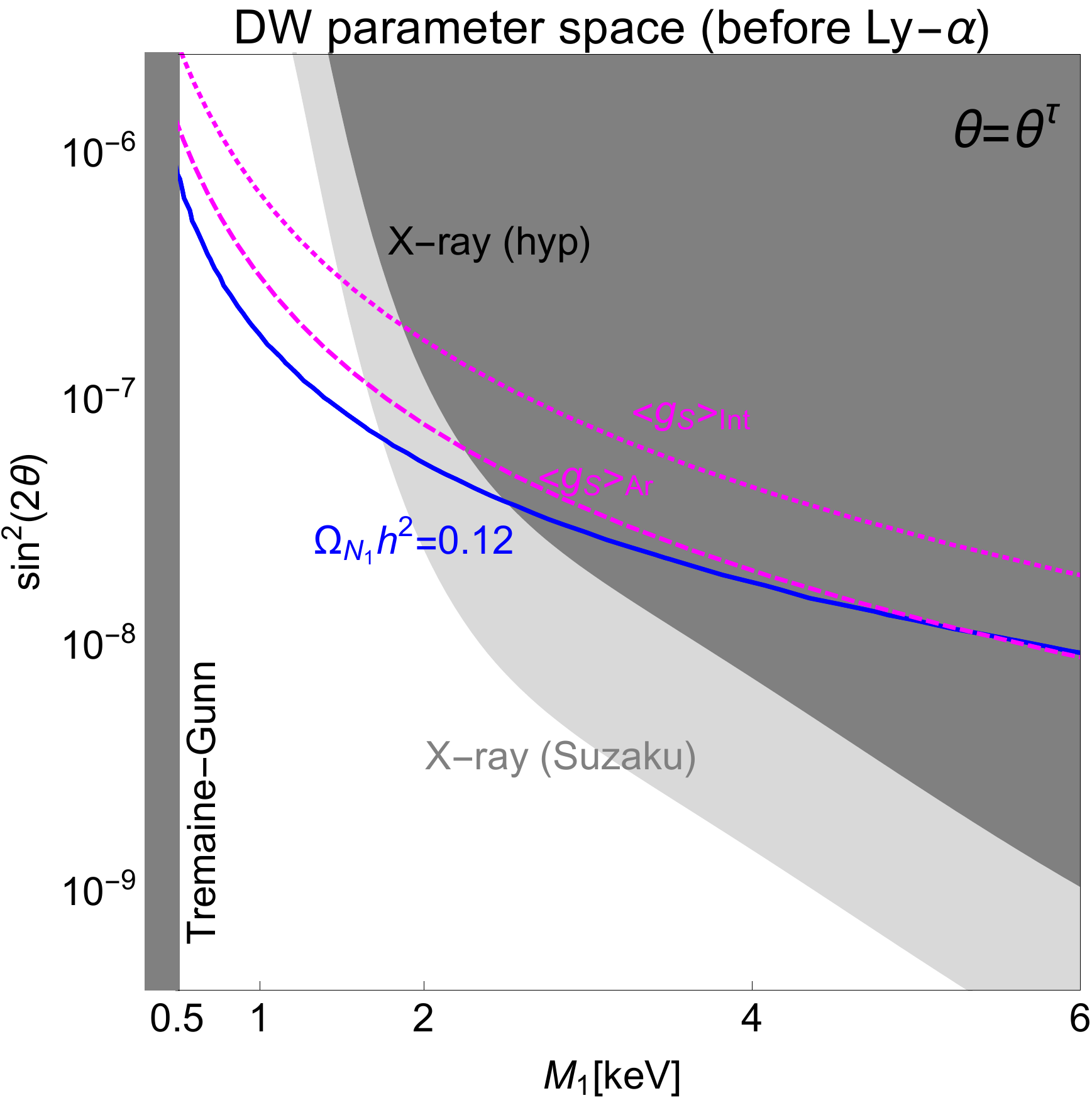}
 \caption{$\theta^\tau=\ssscript{\theta}{max}{hyp}\left(M_1\right)$.}
 \end{subfigure}
 \caption{Parameter space for pure DW production before applying limits from structure formation with isoabundance lines for numerical limits and approximative results.}
 \label{fig:ThetaMPlanePureDWIsoabundance}
\end{figure}

\begin{figure}
 \centering
 \includegraphics[width=0.5 \textwidth]{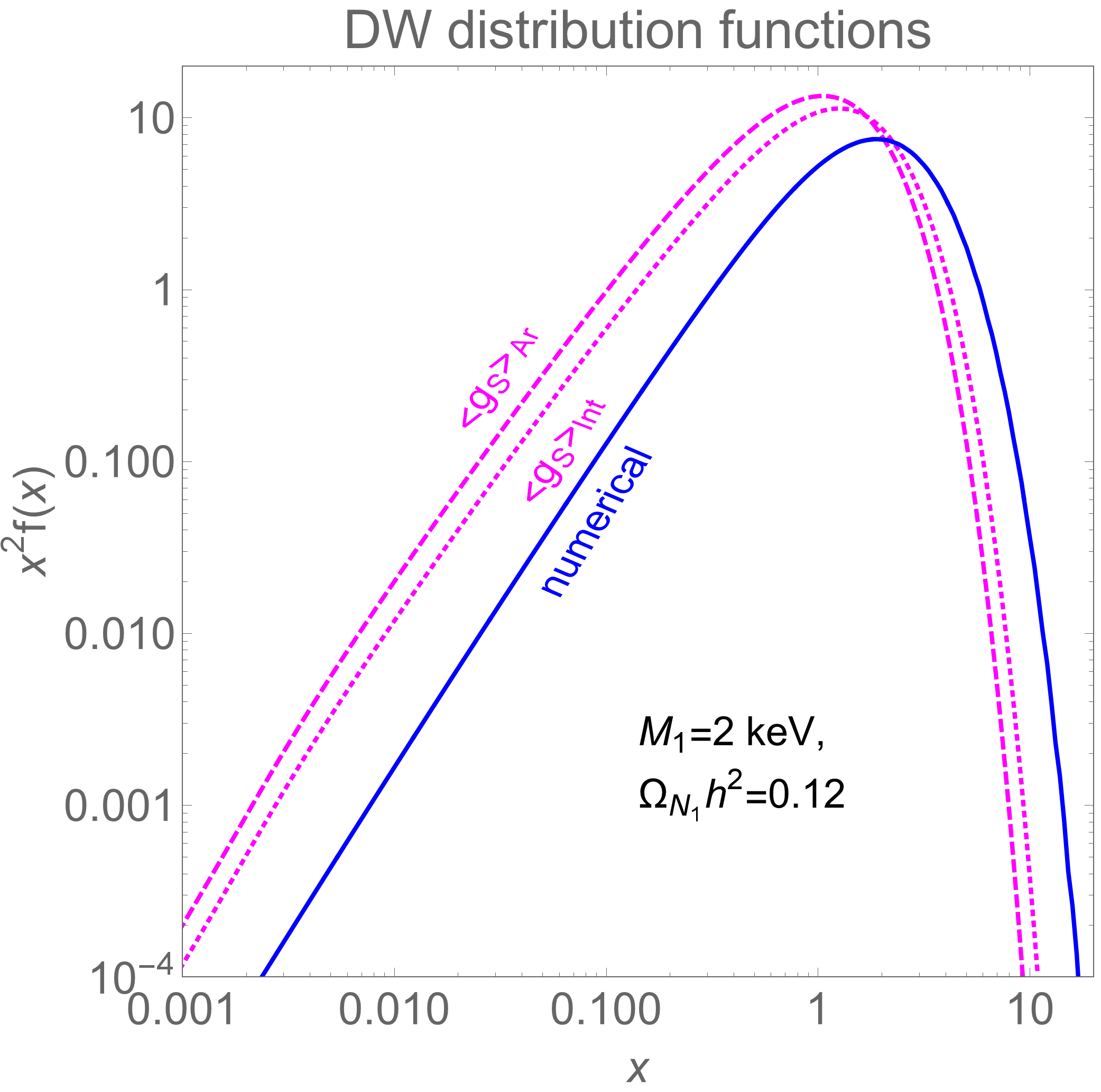}
 \caption{Numerical and approximative distributions for pure electron mixing (with the angle fixed to obtain the relic abundance) and a sterile neutrino mass of $\unit{2}{keV}$.}
 \label{fig:PureDW2keVBenchmark}
\end{figure}

We want to emphasise once more that the best choice of $\av{g_S}$ is \emph{a priori} unclear. To give an impression of the significance of different meaningful choices, we show in \Figref{fig:ThetaMPlanePureDWIsoabundance} numerical and estimated isoabundance lines in the plane spanned by $M_1$ and $\sin^2\left(2\theta\right)$ for the cases of $e$-, $\mu$-, and $\tau$-mixing. The blue curve represents the contour where a pure DW production yields the right abundance if calculated numerically, while the magenta lines use two different \emph{a priori} choices of $\av{g_S}$, namely
\begin{align}
 \av{g_S}_\mathrm{Ar}&\equiv\frac{g_S\left(\sub{T}{ini}\right)-g_S\left(\sub{T}{final}\right)}{2} \; \label{eq:Def:AvGAr} \quad \text{as arithmetic mean and}\\
 \av{g_S}_\mathrm{Int}&\equiv \frac{1}{\sub{T}{ini}-\sub{T}{final}}\int\limits_{\sub{T}{f}}^{\sub{T}{ini}}{\diffd T\ g_S\left(T\right)} \label{eq:Def:AvGInt} \quad \text{as integral mean}\;. 
\end{align}
The figure contains limits from X-ray observations (for a detailed explanation of the most conservative bound dubbed \emph{hyp}, see \Secref{sec:NumericalAnalysisDW}) as well as the Tremaine-Gunn bound. In all three cases, using a meaningful average $\av{g_S}$ can lead to an overestimate of the square of the mixing angle by about half an order of magnitude when fixing the abundance to the current best-fit value from Planck~\cite{Planck:2015xua}.
To complete this discussion, we show the numerical distribution function as compared to the estimated ones in \Figref{fig:PureDW2keVBenchmark}. We have chosen a mass of $M_1=\unit{2}{keV}$ and pure $e$-mixing since, according to \Figref{fig:ThetaMPlanePureDWIsoabundance:e}, this is about the maximum mass that can reproduce the observed relic abundance without violating the most conservative \emph{hyp} X-ray limit.  
Of course we anticipate that all these spectra will not be in agreement with bounds from Ly-$\alpha$ observations. Nonetheless, we will show the effect of only estimating the distribution for the sake of completeness.

Before advancing to more sophisticated analysis methods in \Secref{sec:StructureFormation}, let us here present a simple way to estimate the compatibility of DW production and structure formation. We will derive a translation between the mass of a non-thermally produced DM particle with mass $M_1$ and a thermal relic with mass $m_w$, such that the implications for structure formation are roughly the same. Such a translation can already be found in \cite[Eq.~(12)]{Markovic:2013iza}. While we could not reproduce the exact numerical coefficients of \cite[Eq.~(12)]{Markovic:2013iza}, because the assumptions made to obtain it are not given explicitly in that paper, we present the most relevant steps of the derivation in a fully parametric way, i.e.~keeping all model dependent parameters in the final result.

The rationale behind our translation is the idea that the average velocity of the DM particles at a certain time after production is a good indicator for structure formation. Let us thus first analyse a thermally produced species of mass $m_w$ and $g_w$ internal d.o.f. that has a temperature $T_w$ (which may deviate from the plasma temperature $T=T_\gamma$). Let us assume that this species decouples at a temperature where the background plasma counts $\sub{g}{dec}$ entropy d.o.f.

Calculating the relic abundance for this species, we find the relation:
\begin{align}
 \Omega_w h^2=\frac{45\zeta\left(3\right)}{2\pi^4} s_0 \frac{h^2}{\sub{\rho}{crit}} \frac{g_w}{\sub{g}{dec}} \left(\frac{T_w}{T_\gamma}\right)^3 m_w \times 
\begin{cases}
3/4 \quad \text{for a fermionic species,} \\ 
1 \quad \text{for a bosonic species,}                                                                                                                          \end{cases}
\end{align}
where $h=H_0/\left(\unit{100}{km s^{-1} Mpc^{-1}}\right)$, $\sub{\rho}{crit}$ is the critical density of the Universe, and $s_0$ is today's entropy density. For a thermal species in the mass range of some keV to MeV, we can safely assume that it has become non-relativistic at matter-radiation equality, such that the average velocity at this epoch is given by:
\begin{align}
  \av{v\left(\ssscriptupper{T}{\gamma}{eq}\right)}
 &= \frac{\int \frac{\diffd^3p}{\left(2\pi\right)^3} \frac{p}{m_w} f_w\left(p\right) g_w }{\int \frac{\diffd^3p}{\left(2\pi\right)^3} f_w\left(p\right) g_w} 
 =\frac{\pi^2}{30}\frac{g_w}{n} \frac{T_w^4}{m_w} \times \begin{cases}
                                                          7/8\quad \text{for a fermionic species,} \\
                                                          1 \quad \text{for a bosonic species.}    
                                                         \end{cases}
\label{eq:VielFormulaAbundanceAvTemperature}
\end{align}
Expressing the inverse of the number density by the abundance $\Omega_w h^2$ and mass
$m_w$, we find
\begin{align}
 \av{v\left(\ssscriptupper{T}{\gamma}{eq}\right)} = \left(\Omega_w h^2\right)^{1/3}
 \frac{\sub{g}{dec}^{4/3}g_w^{1/3}}{\sub{g}{eq}} \ssscriptupper{T}{\gamma}{eq} \frac{1}{m_w^{4/3}} \left(\frac{\sub{\rho}{crit}}{s_0 h^2}\right)^{1/3} \frac{\pi^{16/3}}{3\zeta\left(3\right)} \times 
 \begin{cases}
 \frac{7}{270\cdot 5^{1/3}} \quad \text{for a fermionic species,} \\
 \frac{1}{15\cdot 180^{1/3}} \quad \text{for a bosonic species.} 
 \end{cases}          
 \label{eq:VielFormulaAbundanceAvTemperature2}
 \end{align}

In contrast to the species that we have discussed now, we do not assume that the sterile neutrino has any thermal history, i.e., we assume no spectral form for the momentum distribution function. We only assume that the production ceases at some temperature $\sub{T}{prod}$ when there are $\sub{g}{prod}$ entropy d.o.f.~in the plasma. We parametrise the average momentum $\av{p}$ at production temperature by $\av{p}=\alpha \sub{T}{prod}$, where $\alpha$ is usually of $\orderof{1}$.
Then, at equality, the average velocity of the sterile neutrino with mass $M_1$ -- again assuming it is non-relativistic -- is given by:
\begin{align}
 \av{v\left(\ssscriptupper{T}{\gamma}{eq}\right)} = \frac{\alpha\ssscriptupper{T}{\gamma}{eq}}{M_1} \left(\frac{\sub{g}{eq}}{\sub{g}{prod}}\right)^{1/3} \,.
 \label{eq:VielFormulaAbundanceAvTemperatureSN}
\end{align}
Equating \eqref{eq:VielFormulaAbundanceAvTemperature2} and \eqref{eq:VielFormulaAbundanceAvTemperatureSN}, we can solve for $M_1$:
\begin{align}
 M_1 = \left(\Omega_w h^2\right)^{-1/3} m_w^{4/3} \left(\frac{\sub{g}{eq}}{\sub{g}{prod}}\right)^{1/3} \left(\frac{g_w}{\sub{g}{dec}}\right)^{4/3} \frac{3\zeta\left(3\right) \alpha}{\pi^{16/3}} \left(\frac{s_0 h^2}{\sub{\rho}{crit}}\right)^{1/3}\times
 \begin{cases}
  \frac{270 \cdot 5^{1/3}}{7} \quad \text{(fermions),} \\
  15 \cdot 180^{1/3} \quad \text{(bosons).}
 \end{cases}
\end{align}
Inserting normalisation constants motivated by the region of interest for $M_1$ and measurements of the relic abundance yields
\begin{align}
 M_1 = 2.46 \alpha \left(\frac{\Omega_w h^2}{0.1225}\right)^{-1/3} \times \left(\frac{m_w}{\unit{1}{keV}}\right)^{4/3} \times \left(\frac{\sub{g}{eq}}{\sub{g}{dec}}\right)^{4/3} \times \left(\frac{g_w}{\sub{g}{prod}}\right)^{1/3}
\end{align}
for a fermionic species. Choosing suitable values for the d.o.f.~and for $\alpha$, we find a result that is in the same ballpark as \cite[Eq.~(12)]{Markovic:2013iza}, but tends to yield a slightly smaller sterile neutrino mass for a fixed thermal mass $m_w$. Also the recent conversion formula between thermal masses and sterile neutrino masses in~\cite{Bozek:2015bdo} yields slightly smaller masses than~\cite[Eq.~(12)]{Markovic:2013iza}. This tendency shows that the conversion formula in \cite[Eq.~(12)]{Markovic:2013iza} is slightly too aggressive, resulting in somewhat too tight limits as we will show in \Secref{sec:StructureFormation}, together with the analysis of the approximations for a pure DW spectrum presented in Eqs.~\eqref{eq:ThermalApprox} -- \eqref{eq:Def:AvGInt}. On the other hand, we will also show in this section that using the exact spectrum could even improve the limits.

%%%%%%%%%%%%%%%%%%%%%%%%%%%%%%%%%%%%%%%%%%%%%%%%%%%%%%%%%%%%%%%%%%%%%%%%%%%%%%%%%%%%%%%%%%%%%%%%%%
\section{\label{sec:NumericalAnalysisDW}Quantitative analysis of the mechanism}
%%%%%%%%%%%%%%%%%%%%%%%%%%%%%%%%%%%%%%%%%%%%%%%%%%%%%%%%%%%%%%%%%%%%%%%%%%%%%%%%%%%%%%%%%%%%%%%%%%

\begin{figure}[t]
 \centering
 \begin{subfigure}{0.32 \textwidth}
 \includegraphics[width=\textwidth]{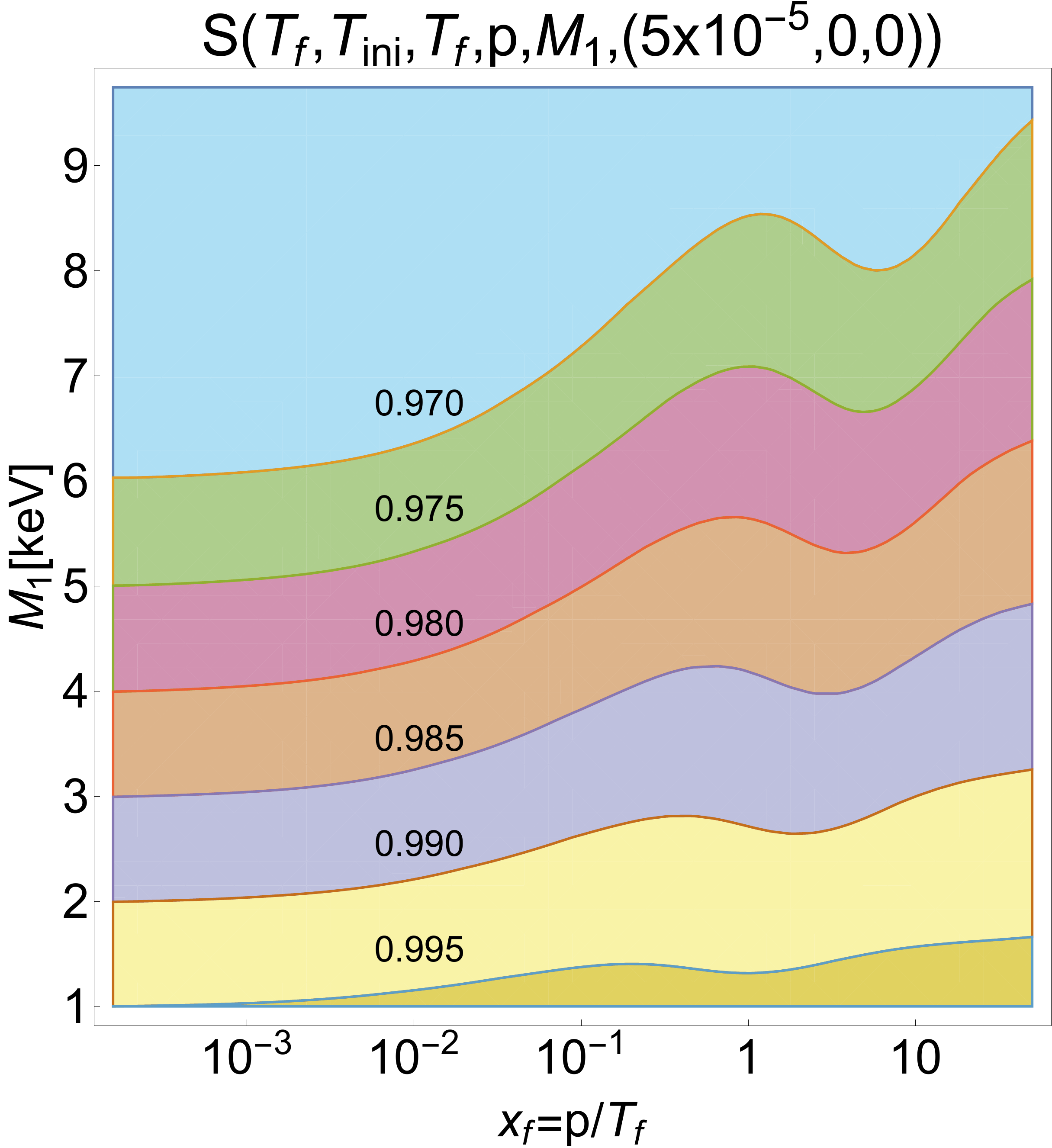}
 \caption{Mixing with $\nu_e$.}
 \end{subfigure}
 \begin{subfigure}{0.32 \textwidth}
 \includegraphics[width=\textwidth]{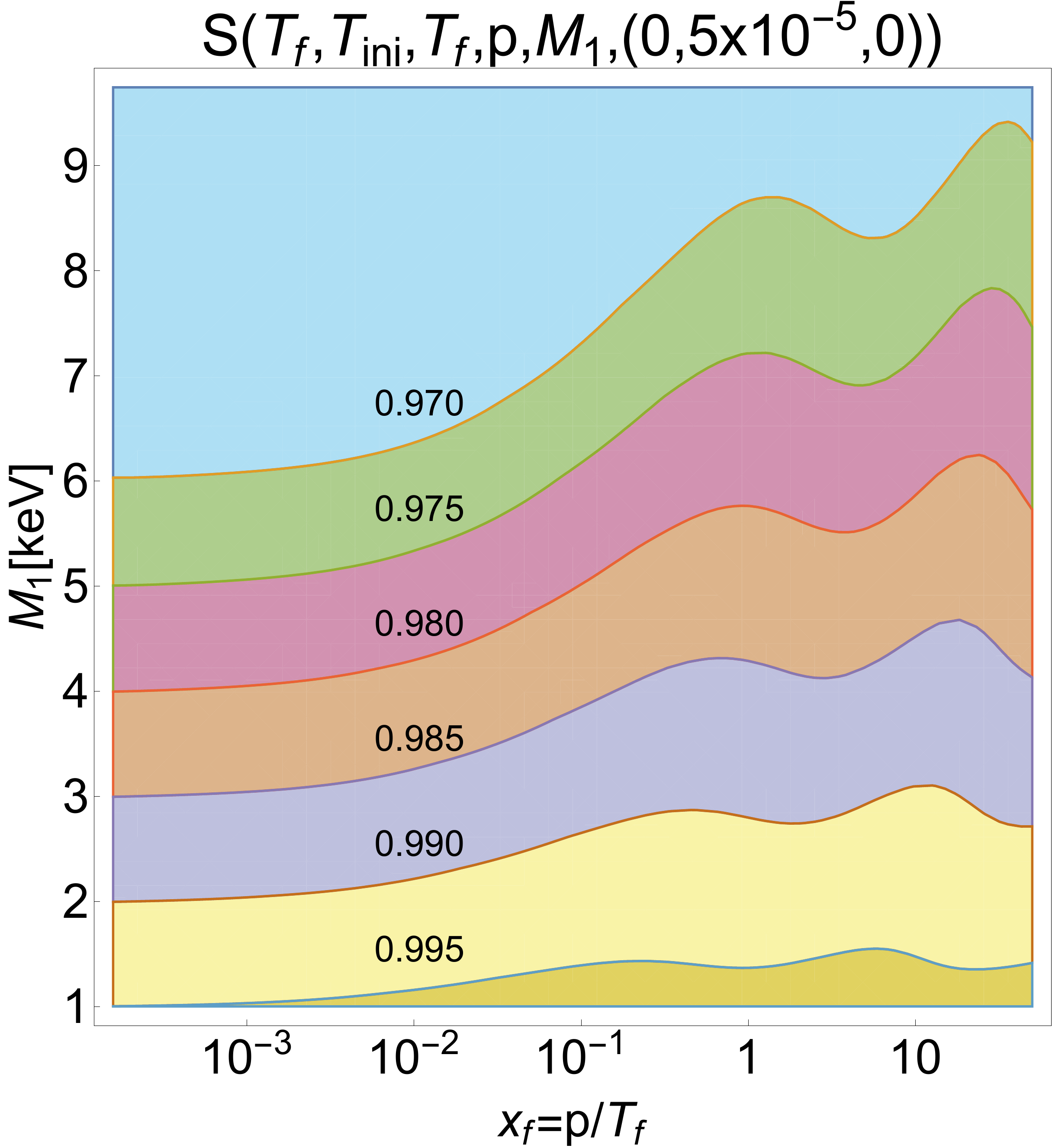}
 \caption{Mixing with $\nu_\mu$.}
 \end{subfigure}
 \begin{subfigure}{0.32 \textwidth}
 \includegraphics[width=\textwidth]{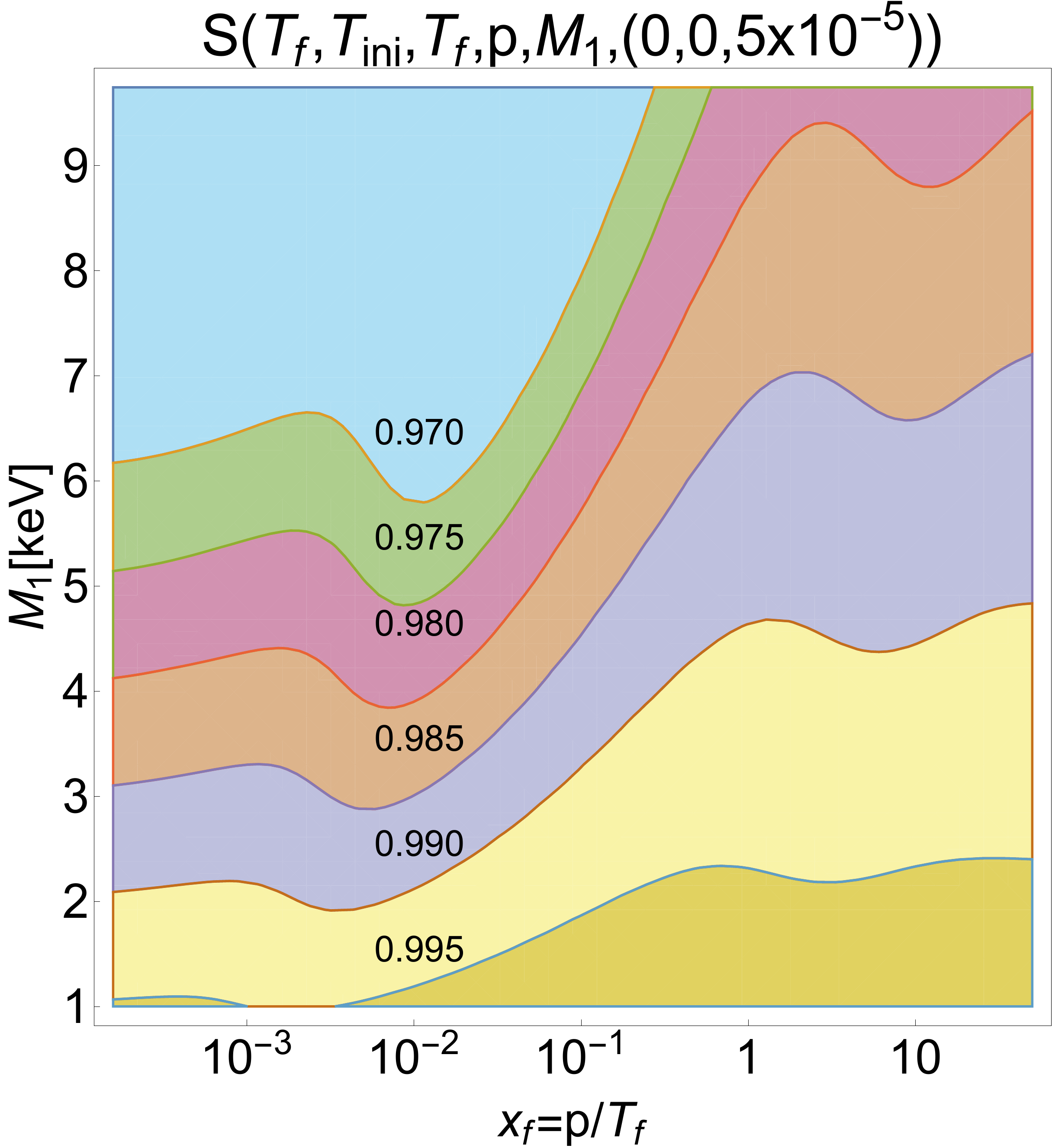}
 \caption{Mixing with $\nu_\tau$.}
 \end{subfigure}
 \caption{\label{fig:SuppressionFactorConstantAngle}Suppression factor $\mathcal{S}\left(\sub{T}{f},\sub{T}{ini},\sub{T}{f},p,M_1,\left(\theta^e,\theta^\mu,\theta^\tau\right)\right)$ in the plane spanned by $\sub{x}{f}$ and $M_1$ for $\sub{T}{ini}=\unit{10}{GeV}$ and $\sub{T}{f}=\unit{10}{MeV}$.}
\end{figure}

\begin{figure}[t]
 \centering
 \begin{subfigure}{0.32 \textwidth}
 \includegraphics[width=\textwidth]{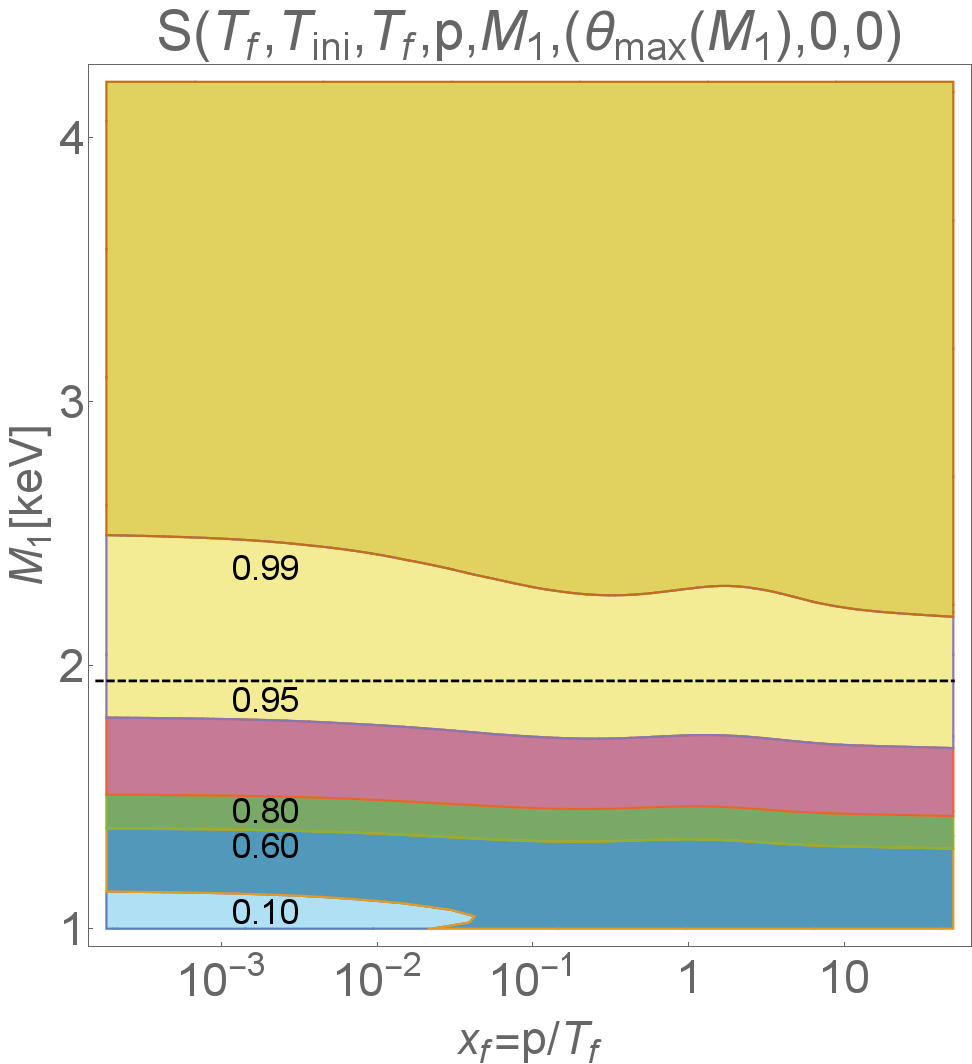}
 \caption{$\theta^e=\ssscript{\theta}{max}{Suzaku}\left(M_1\right)$.}
 \end{subfigure}
 \begin{subfigure}{0.32 \textwidth}
 \includegraphics[width=\textwidth]{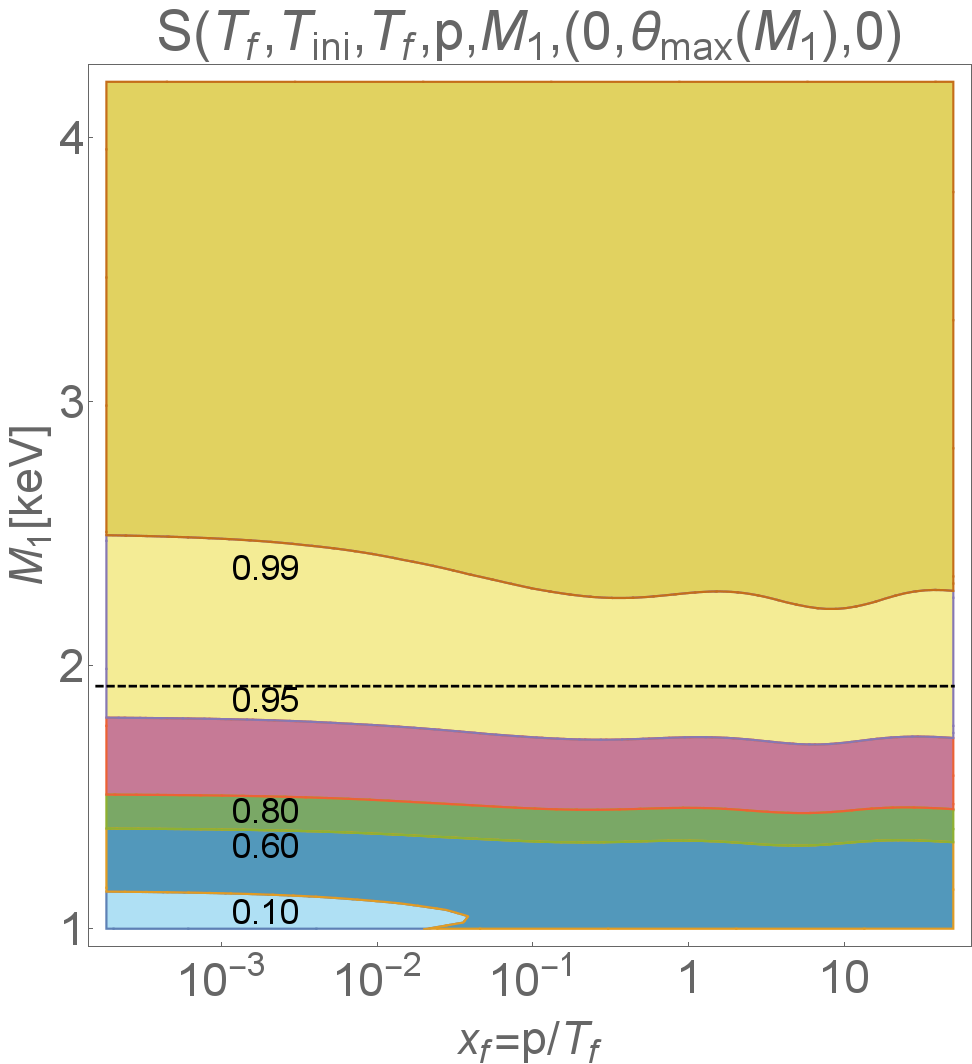}
 \caption{$\theta^\mu=\ssscript{\theta}{max}{Suzaku}\left(M_1\right)$.}
 \end{subfigure}
 \begin{subfigure}{0.32 \textwidth}
 \includegraphics[width=\textwidth]{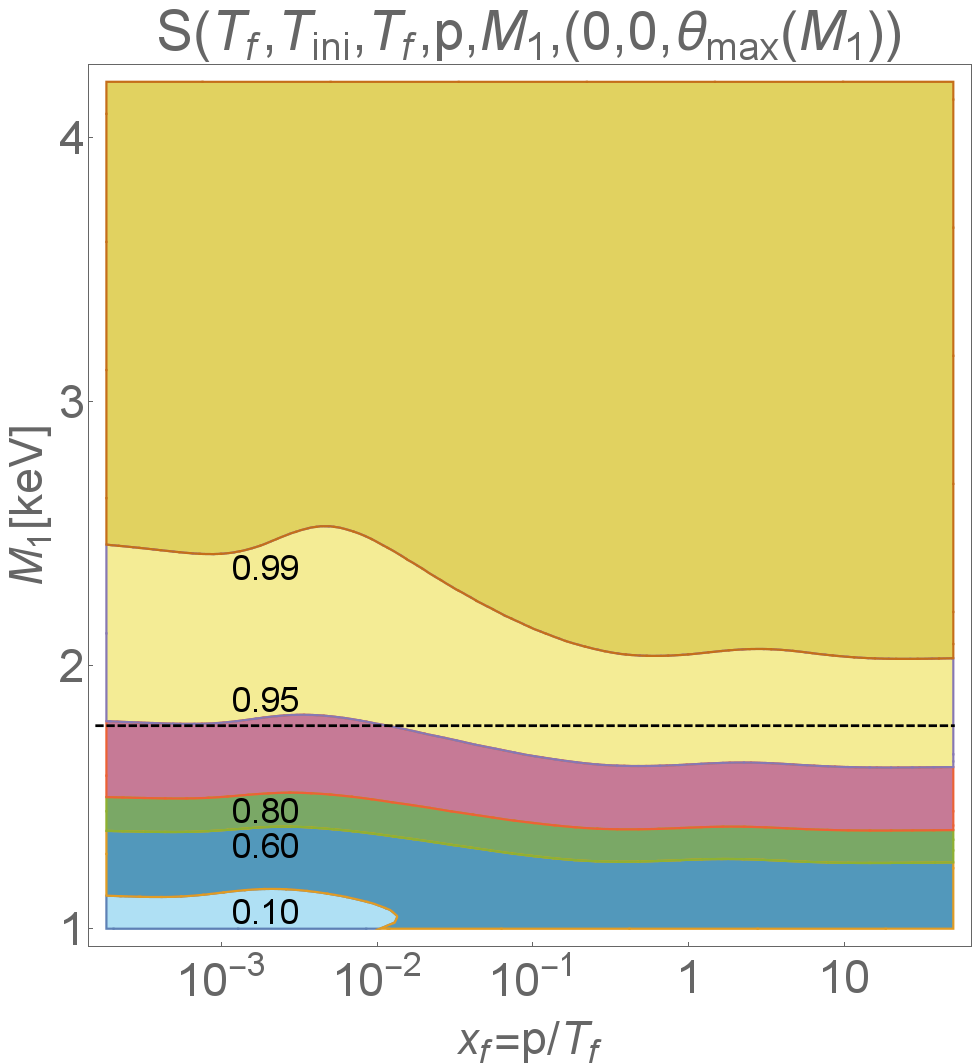}
 \caption{$\theta^\tau=\ssscript{\theta}{max}{Suzaku}\left(M_1\right)$.}
 \end{subfigure}
 \\
 \begin{subfigure}{0.32 \textwidth}
 \includegraphics[width=\textwidth]{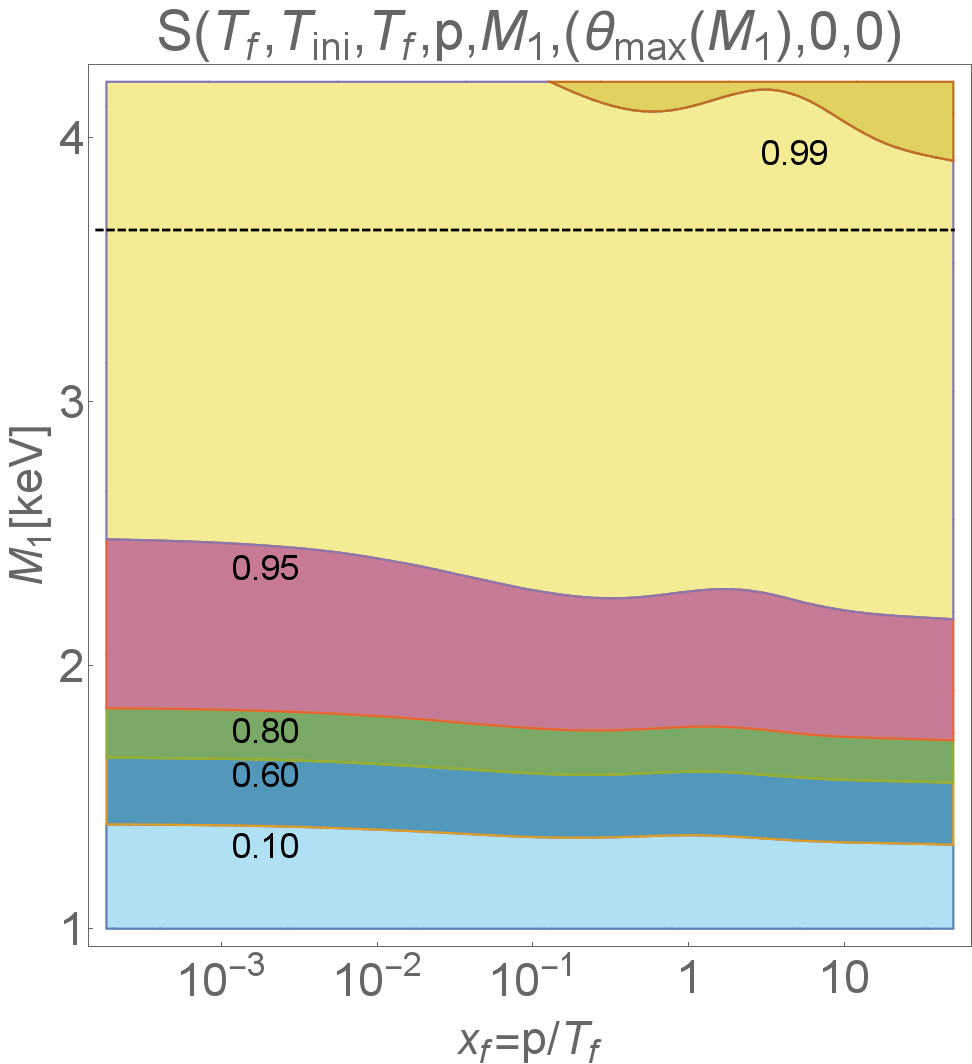}
 \caption{$\theta^e=\ssscript{\theta}{max}{hyp}\left(M_1\right)$.}
 \end{subfigure}
 \begin{subfigure}{0.32 \textwidth}
 \includegraphics[width=\textwidth]{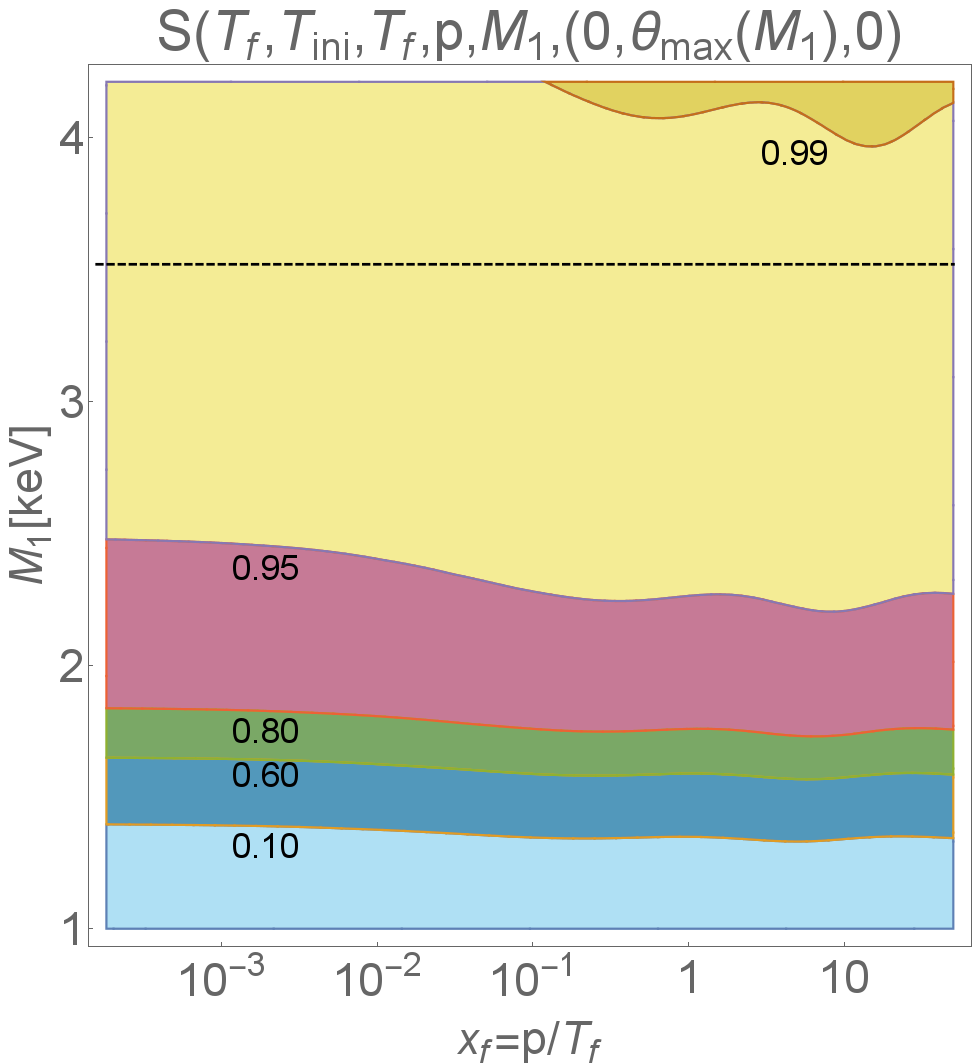}
 \caption{$\theta^\mu=\ssscript{\theta}{max}{hyp}\left(M_1\right)$.}
 \end{subfigure}
 \begin{subfigure}{0.32 \textwidth}
 \includegraphics[width=\textwidth]{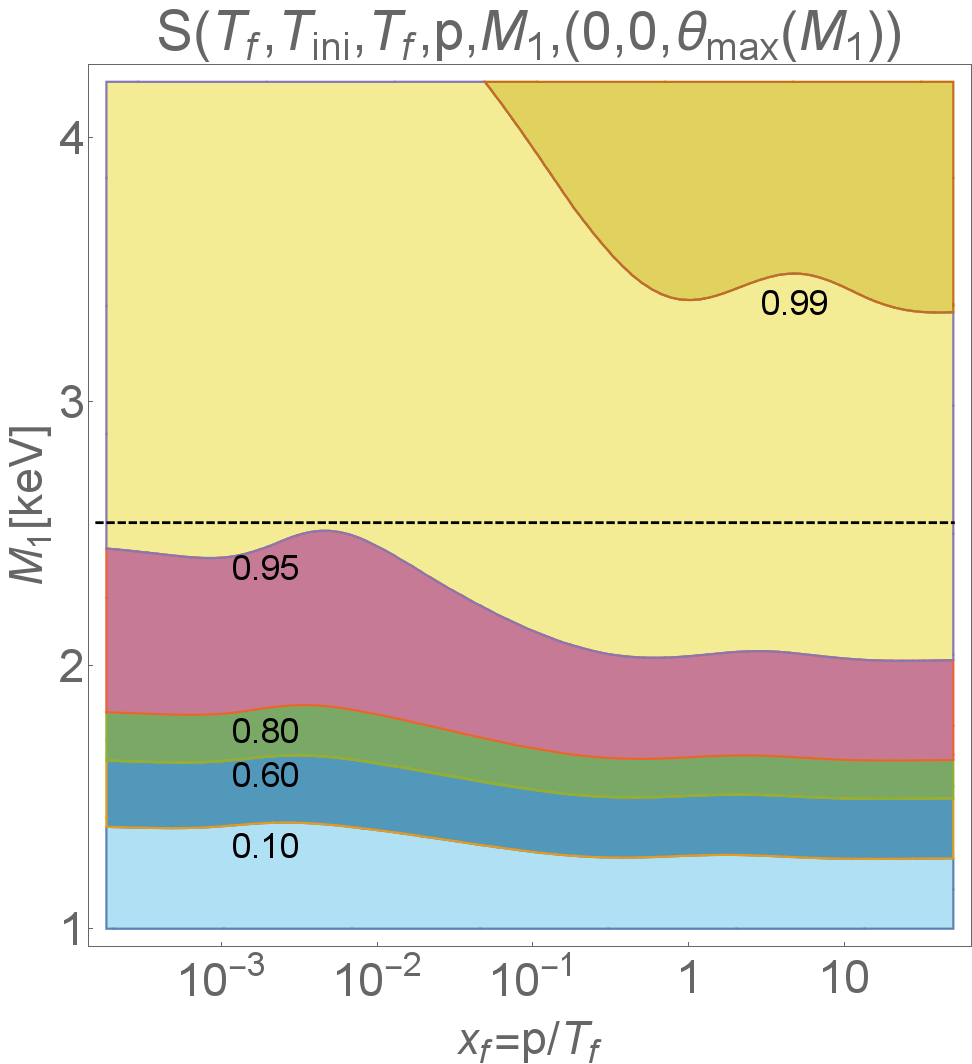}
 \caption{$\theta^\tau=\ssscript{\theta}{max}{hyp}\left(M_1\right)$.}
 \end{subfigure}
 \caption{\label{fig:SuppressionFactorMaximalAngle}Contours of $\mathcal{S}$ in the $\sub{x}{f}$-$M_1$ plane for $\theta=\sub{\theta}{max}$ in two cases (see text for details). We assume pure $e$ ($\mu$) [$\tau$] mixing in the left (centre) [right] columns. Dashed lines indicate the mass below which maximal mixing leads to overclosure for pure DW production.}
\end{figure}

Having seen the formal-analytical solution of \equref{eq:BoltzmannDWAbstract}, let us have a look at the numerical properties of this solution, using numerical fits for $g_S$ presented in~\cite{Wantz:2009it}. We start by discussing whether or not it is a good approximation to just add the DW-component by hand to any (correctly redshifted) initial distribution of sterile neutrinos, i.e.,~to neglect potential damping or distortion effects of the active-sterile conversion on the spectrum. This has been done in several references~\cite{Abada:2014zra,Merle:2014xpa,Kang:2014cia,Humbert:2015epa}, however, in \emph{none} of those works the reliability of this assumption has  actually been checked. Given our general solution, we will now show that the approximation applied in the earlier references does indeed work. We thus provide an \emph{a-posteriori justification} of the treatments used in previous works.

Looking again at the formal solution reported in \equref{eq:SolutionExactCondensed}, we realise that this question about the reliability of the approximation described above can easily be answered by analysing how much the factor $\mathcal{S}\left(\sub{T}{f},\sub{T}{ini},\sub{T}{f},p\right)$ can deviate from unity. To this end, we plot the contours of $\mathcal{S}$ in the plane spanned by $x_{\rm f}=p/\sub{T}{f}$ and $M_1$, both for a (quite large) example value of $5\cdot 10^{-5}$ for the different mixing angles $\theta^{e,\mu,\tau}$ (\Figref{fig:SuppressionFactorConstantAngle}) and for the maximal mixing angle allowed by X-ray observations dubbed $\theta=\sub{\theta}{max}\left(M_1\right)$, see \Figref{fig:SuppressionFactorMaximalAngle}. We chose $\sub{T}{ini}=\unit{10}{GeV}$ and $\sub{T}{f}=\unit{10}{MeV}$, thereby spanning the range relevant for DW production~\cite{Abazajian:2001nj}. Extending this temperature range does not alter the results.

In the case of $\theta=\sub{\theta}{max}$, we show the results of the analysis for two different versions of the X-ray bounds. The top panel uses an analysis based on data from Suzaku~\cite{Sekiya:2015jsa} (dubbed \emph{Suzaku}), which update the combined limits obtained in~\cite{Merle:2013ibc}, while the bottom panel relaxes the limits on $\sub{\theta}{max}^2$ by a factor of $5$ (dubbed \emph{hyp}). The latter option illustrates what would change if our current face value bounds were in fact too strong, which is not completely unrealistic given the intrinsic uncertainties of X-ray observations.\footnote{The reasons for this is not our astrophysics colleagues delivering a bad work, but instead it is intrinsically difficult to determine the active-sterile mixing angle because it ultimately requires a measurement of the signal \emph{intensity}.} We further indicate by a dashed line the mass below which the maximally allowed mixing angles overproduces sterile neutrinos from DW alone. This means that all masses \emph{below} this line are excluded, \emph{irrespective of the initial distribution before onset of DW production.}

As evident from \Figref{fig:SuppressionFactorMaximalAngle}, the maximal suppression of any momentum mode in the initial spectrum can be a few percent at most. No matter which scenario is taken, it is either the X-ray bound itself which forces the mixing angles to be small and thus implies a small effect (for sterile neutrino masses of roughly 4~keV and larger), or it is the constraint of not producing too much DM which does not allow for large active-sterile mixing (below a mass of, say, 2~keV). Only around sterile neutrino masses of $M_1 \sim 3$~keV the DW modification could possibly be significant. Even in that region, however, the effect turns out to be very marginal, and it never amounts to more than $5\%$.

This shows that \emph{simply adding a DW component to any previously produced sterile neutrino population is indeed a pretty good approximation}. No big modifications of the spectrum are expected. Comparing \Figref{fig:SuppressionFactorConstantAngle} to \Figref{fig:SuppressionFactorMaximalAngle}, we can make the following observation: for a fixed mixing angle, the suppression becomes stronger for larger $M_1$, while small values of $M_1$ produce the highest suppression if the mixing angle is fixed to its upper limit value. This reversing of the dependence of $\mathcal{S}$ on $M_1$ can be understood by the fact that decay rate of sterile neutrinos to photons and ordinary neutrinos scales as $M_1^5$. Therefore, the X-ray bounds for small neutrino masses are \emph{much} less stringent than for higher masses, thereby overcompensating the decrease of $\mathcal{S}$ for decreasing $M_1$ and fixed mixing angle.

Note that, while the DW modification turns out to be negligible in practice, this result was \emph{not} a priori clear beforehand. It can however be seen in our general solution: glancing at Eq.~\eqref{eq:Def:S} and at Fig.~\ref{fig:h-variance}, it is evident that the function $h$ is very small, thereby implying a nearly vanishing exponent and thus a factor of $\mathcal{S}$ close to unity. Physically, this comes simply from $h$ being proportional to the tiny mixing angle square, cf.\ Ref.~\cite{Abazajian:2001nj}. However, given that $h$ is a dimensionful quantity, one cannot easily associate the label \emph{small} to it, such that it indeed would have been impossible to conclude about the actual value of $\mathcal{S}$ just from the fact that some small quantity is involved in $h$.

%%%%%%%%%%%%%%%%%%%%%%%%%%%%%%%%%%%%%%%%%%%%%%%%%%%%%%%%%%%%%%%%%%%%%%%%%%%%%%%%%%%%%%%%%%%%%%%%%%
\section{\label{sec:StructureFormation}Constraints from structure formation}
%%%%%%%%%%%%%%%%%%%%%%%%%%%%%%%%%%%%%%%%%%%%%%%%%%%%%%%%%%%%%%%%%%%%%%%%%%%%%%%%%%%%%%%%%%%%%%%%%%
It is well known that structure formation imposes stringent constraints on the minimal sterile neutrino mass produced by the DW mechanism. In combination with the bounds from X-ray data, they rule out DW sterile neutrinos as the dominant DM component~\cite{Seljak:2006qw,Horiuchi:2013noa}. Different sterile neutrino production mechanisms may still have subdominant DW contributions, though, as explained in the sections above. It is therefore useful to identify the maximum allowed contribution of DW sterile neutrinos with respect to the total DM budget.

To constrain the DW sterile neutrino abundance, we assume the limiting case of a mixed DM model where the remaining DM component is considered to be perfectly cold. The structure formation of such a scenario has been investigated in various other studies (see e.g.~\cite{Maccio':2012uh,Anderhalden:2012jc}) and there are constraints based on both Lyman-$\alpha$ forest~\cite{Boyarsky:2008mt} and on dwarf galaxy counts~\cite{Schneider:2014rda}. In practice, the remaining DM component does not have to be perfectly cold (as for example in the case of resonantly produced sterile neutrino DM where the colder component still consists of lukewarm DM~\cite{Boyarsky:2008mt}), which would lead to even stronger constraints on the abundance of the DW component.

In this paper we apply the method presented in Ref.~\cite{Schneider:2014rda}, which consists of comparing the expected number of sub-haloes to the observed abundance of satellite galaxies in the Milky-Way. A good estimate of the sub-halo abundance is given by the relation:
\begin{equation}\label{Nsub}
\frac{{\rm d}N_{\rm sh}}{{\rm d} \ln M_{\rm sh}}=\frac{1}{44.5}\frac{1}{6\pi^2} \left(\frac{M_{\rm hh}}{M_{\rm sh}}\right)\frac{P(1/R_{\rm sh})}{R_{\rm sh}^3\sqrt{2\pi(S_{\rm sh}-S_{\rm hh})}} \,,
\end{equation}
where variances $S_i$ and masses $M_i$ are defined as
\begin{equation}
S_{i}=\frac{1}{2\pi^2}\int_0^{1/R_i}{\rm d}k\ k^2P(k),\hspace{0.5cm}M_i= \frac{4\pi}{3}\Omega_{\rm m} \rho_c(2.5R_i)^3,\hspace{1cm}i=\{\rm sh,hh\} \,,
\end{equation}
with sh and hh standing for {\it subhalo} and {\it host-halo}, respectively. Eq.~\eqref{Nsub} is based on a modification of the extended Press-Schechter recipe (using a sharp-$k$ window function) described in~\cite{Schneider:2013ria} (see also~\cite{Benson:2012su}) and a normalisation to $N$-body simulations of~\cite{Lovell:2013ola} to account for tidal stripping. The only input it requires is the linear power spectrum $P\left(k\right)$ for a given cosmology and DM model.

For the estimate of Milky-Way satellite numbers, we follow the method of~\cite{Polisensky:2010rw} (see also~\cite{Kennedy:2013uta}) which consists of adding up 11 classical and 52 ultra-faint dwarf galaxies (the latter is an estimate based on the 15 ultra-faint dwarfs observed by SDSS multiplied by a factor of 3.5 to account for the limited SDSS sky coverage). We refer to the Refs.~\cite{Polisensky:2010rw,Kennedy:2013uta,Schneider:2014rda} for details about this method including discussions about statistical and systematical uncertainties.

Based on stellar kinematics it is possible to estimate the total mass of the observed Milky-Way satellites, albeit with rather large systematic uncertainties. We therefore only assume that all known satellites have total halo masses above $10^8 h^{-1} M_{\odot}$, which is a conservative estimate based on results from  Ref.~\cite{Brooks:2012vi}. For the mass of the Milky-Way, we consider the range $5.5\times10^{11}<M_{\rm hh}<3.2\times10^{12} h^{-1} M_{\odot}$ given by Ref.~\cite{Guo:2009fn}.

\begin{figure}[t]
 \centering
    \includegraphics[width=1.0\textwidth]{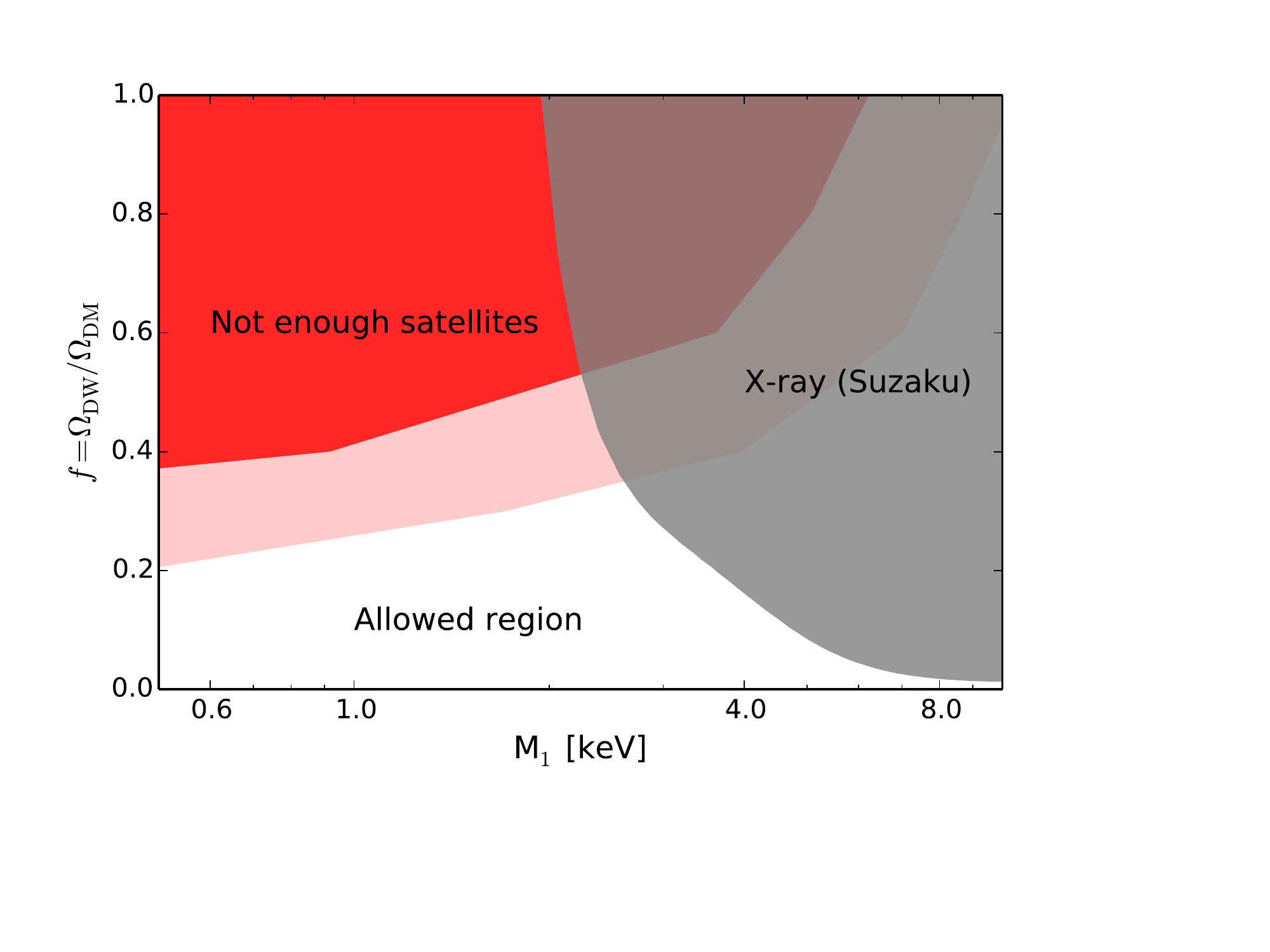}
  \caption{\label{fig:DWabundance}Constraints from Milky-Way satellite counts and X-ray emission (Suzaku) on the fraction of DM produced by the DW mechanism. The pink and red areas represent bounds assuming a Milky-Way mass of 1.2 and 3.2 $\times 10^{12} M_{\odot}/h$ (corresponding to the average and a maximum mass estimate by Ref.~\cite{Guo:2009fn}).}
\end{figure}

The bounds on the maximally allowed abundance of DW sterile neutrinos are shown in Fig.~\ref{fig:DWabundance}. Assuming an average Milky-Way mass of $1.2\times 10^{12} M_{\odot}/h$ (pink exclusion area) the allowed fraction of DW sterile neutrinos (i.e.~$f=\Omega_{\rm DW}/\Omega_{\rm DM}$) never exceeds $f\sim0.3$. It is strongly suppressed beyond $M_1=3$ keV due to the X-ray limit from Suzaku. The allowed fraction increases to a maximum of $f=0.5$ if a very large Milky-Way mass of $3.2\times 10^{12} M_{\odot}/h$ is assumed instead (red exclusion area). Note that our analysis rules out a sizable fraction of parameter space that was found to be valid in a similar analysis done in Ref.~\cite{Abada:2014zra}. Especially in the phenomenologically interesting region around $\unit{7}{keV}$, the maximal fraction of sterile neutrinos produced from the DW mechanism in Ref.~\cite{Abada:2014zra} turns out to be far too optimistic and is not in agreement with our results using the new Suzaku data, as to be expected from the tight constraints on the mixing angle for such masses.

\begin{figure}
	\centering
	\includegraphics[width=0.9 \textwidth]{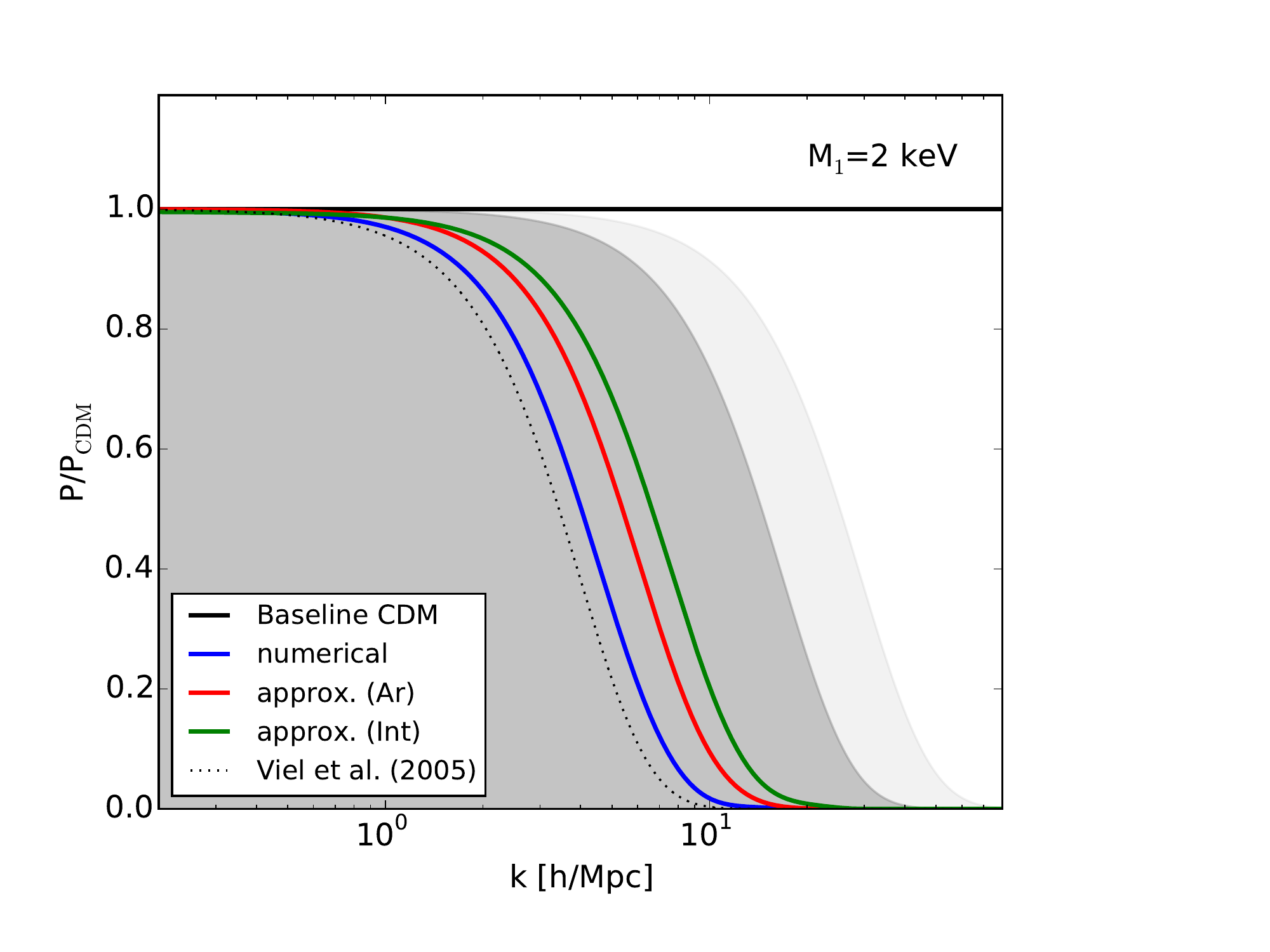}
	\caption{Relative power spectra for sterile neutrinos of a mass of $M_1=\unit{2}{keV}$ being produced by the DW mechanism exclusively. Using the exact numerical version of the MDF will yield the tightest constraints on a model with a non-negligible fraction of sterile neutrinos produced via the DW mechanism. The light- and dark-grey shaded areas illustrate the 2$\sigma$ exclusion limits from Lyman-$\alpha$ observations~\cite{Viel:2013apy} and dwarf galaxy counts~\cite{Polisensky:2010rw}.}
	\label{fig:PkpureDWApproxNum}
\end{figure}

As announced in \Secref{sec:DWInitialAbundance:StandardDW}, we complete this session by showing the relative power spectra computed from the different versions of the MDF in \Figref{fig:PkpureDWApproxNum}. The blue solid line (to the very left) describes the full numerical solution, while the red and the green lines correspond to the different choices of $\av{g_S}$ in the approximation of \equref{eq:ThermalApprox}. The dotted black line is the approximation used in~\cite{Viel:2005qj}. As already discussed, in the case of the two approximations, the relative power spectra fall off only at higher wave-numbers as compared to the numerical case and are therefore too conservative. On the other hand side, the approximation used in Ref.~\cite{Viel:2005qj} is slightly too restrictive, showing that the correct treatment of a DW component in a mixed model for sterile neutrino DM might re-allow formerly borderline cases.

%%%%%%%%%%%%%%%%%%%%%%%%%%%%%%%%%%%%%%%%%%%%%%%%%%%%%%%%%%%%%%%%%%%%%%%%%%%%%%%%%%%%%%%%%%%%%%%%%%
\section{\label{sec:SDCase}Example: Initial abundance from scalar decay}
%%%%%%%%%%%%%%%%%%%%%%%%%%%%%%%%%%%%%%%%%%%%%%%%%%%%%%%%%%%%%%%%%%%%%%%%%%%%%%%%%%%%%%%%%%%%%%%%%%

Having laid the technical foundations for the computation of DW production combined with an initial abundance, let us now proceed to a well-motivated mechanism that could produce such an initial spectrum: adding one real scalar singlet $S$, coupled to the Higgs via some portal coupling $\lambda$, and one sterile neutrino $N_1$, coupled to the scalar via a Yukawa-type interaction with coupling strength $y$, can do this job. This setting was extensively discussed in the literature (see e.g.~\cite{Petraki:2007gq, Merle:2013wta,Adulpravitchai:2014xna}). A detailed analysis on the level of distribution functions was presented in an earlier work by two of us~\cite{Merle:2015oja}, however, with the assumption of zero active-sterile mixing. We now use the opportunity to kill two birds with one stone: the shortcoming of the previous paper will be rectified simultaneously to delivering an explicit example for the current paper.

In order not to unnecessarily prolong this text, we will only sketch the most important aspects of scalar decay production. We refer to Ref.~\cite{Merle:2015oja} for details.

%%%%%%%%%%%%%%%%%%%%%%%%%%%%%%%%%%%%%%%%%%%%%%%%%%%%%%%%%%%%%%%%%%%%%%%%%%%%%%%%%%%%%%%%%%%%%%%%%%
\subsection{\label{sec:SDCase:Model} Singlet scalar decay production of sterile neutrinos}
%%%%%%%%%%%%%%%%%%%%%%%%%%%%%%%%%%%%%%%%%%%%%%%%%%%%%%%%%%%%%%%%%%%%%%%%%%%%%%%%%%%%%%%%%%%%%%%%%%

Our setup extends the SM by two gauge singlets, one real scalar $S$ and one sterile neutrino $N_1$. The Lagrangian is given by:
\begin{equation}
 \Lag = \LagArg{SM} + \left[i\overline{N_1} \delslashed N_1 + \frac{1}{2}\left(\partial_\mu S\right)\left(\partial^\mu S\right) - \frac{y}{2}S\overline{N_1^c}N_1 +\hc \right] - \sub{V}{scalar} + \Lag_\nu \,,
 \label{eq:ModelLagrangian}
\end{equation}
where $\LagArg{SM}$ is the SM Lagrangian, $\Lag_\nu$ is the part of the Lagrangian that gives mass to the active neutrinos [and contains the mixing angles $\left(\theta^e,\theta^\mu,\theta^\tau\right)$], and $\sub{V}{scalar}$ is the scalar potential containing a Higgs portal coupling $\propto \lambda \left(H^\dagger H\right)S^2$ between the SM-like Higgs $H$ and the new scalar. With this Lagrangian, two basic processes are relevant for the production of sterile neutrinos:
\begin{align*}
 hh &\leftrightarrow SS \;, \\
 S &\rightarrow N_1N_1 \,.
\end{align*}

As in~\cite{Merle:2015oja}, we will apply the assumption that the scalar decay occurs well before the QCD transition, such that $g_S\sim \mathcal{O}(100)$ can be taken to be roughly constant during DM production. Denoting the scalar mass as $m_S$, we can express our results as functions of two effective parameters~\cite{Merle:2015oja},
\begin{equation}
 \left\{
 \begin{array}{ll}
 \textrm{\emph{the effective decay width}:} & \CGamma \equiv \frac{M_0}{m_S}\frac{\Gamma}{m_S}\,, \ \ \ \text{and} \hfill \hfill \hfill\\
 \textrm{\emph{the effective (squared) Higgs portal}:    } & \CHP \equiv \frac{M_0}{m_S} \frac{\lambda^2}{16\pi^3} \,.
 \end{array}\right.
 \label{eq:DefCGammaAndCHP}
\end{equation}

A few brief remarks on these quantities are in order (cf.~\cite[Secs.~4 and 6]{Merle:2015oja}):
\begin{itemize}

 \item The effective Higgs portal $\CHP$ determines whether or not the scalar equilibrates in the early Universe. If $\CHP$ is small enough, the scalar undergoes a \emph{freeze-in process}~\cite{Merle:2013wta} before decaying into sterile neutrinos, while it instead \emph{freezes out} if $\CHP$ is large enough~\cite{Petraki:2007gq}. This translates directly into different sterile neutrino MDFs.
 
 \item The effective decay width $\CGamma$ determines how fast the scalar decays. It thus dictates \emph{when} the energy stored in the mass of the scalar is injected into the Universe in the form of highly relativistic sterile neutrinos (recall that $M_1\ll m_S$). Therefore it ultimately decides about the spectrum being warmer or cooler, and it is very relevant to determine the consequences for structure formation. Also, if the scalar equilibrates, there is an upper limit for $\CGamma$ beyond which the particle number of sterile neutrinos becomes so large that any mass not violating the Tremaine-Gunn bound~\cite{Tremaine:1979we} will always overclose the Universe, see Ref.~\cite{Merle:2015oja} for details.
\end{itemize}

%%%%%%%%%%%%%%%%%%%%%%%%%%%%%%%%%%%%%%%%%%%%%%%%%%%%%%%%%%%%%%%%%%%%%%%%%%%%%%%%%%%%%%%%%%%%%%%%%%
\subsection{\label{sec:SDCase:Examples} Benchmark cases with and without a DW component}
%%%%%%%%%%%%%%%%%%%%%%%%%%%%%%%%%%%%%%%%%%%%%%%%%%%%%%%%%%%%%%%%%%%%%%%%%%%%%%%%%%%%%%%%%%%%%%%%%%

With the preceding discussion we can motivate our choice of three benchmark cases that will be taken as initial abundance for the DW production: \Figref{fig:SuppressionFactorMaximalAngle} suggests that the sterile neutrino mass $M_1$ should not exceed a few $\unitonly{keV}$ in order not to violate the overabundance bound while simultaneously not having very small mixing angles due to X-ray constraints. First estimates on structure formation simultaneously demand a very early production for sterile neutrinos in this mass range (see \cite[Fig.~8]{Merle:2015oja}). Recall that a very early production is equivalent to a large $\CGamma$, which restricts us to the \emph{freeze-in} region of the scalar decay parameter space. The exact parameters defining these cases will be summarised in \Tabref{tab:OverviewBenchmarkCases}. Each case will in turn be subdivided into three subcases. The first subcase -- dubbed $\alpha$ -- corresponds to a fixed choice of $\left(\CHP, \CGamma, \left(\theta^e,\theta^\mu,\theta^\tau\right)=\left(0,0,0\right)\right)$, i.e., DW production switched off completely.

Since we assumed $M_1\ll m_S$, the mass of the sterile neutrino is completely irrelevant when calculating the distribution function $f_N\left(p, T\right)$ from scalar decay. Combining this fact with the parametrisation from \equref{eq:DefCGammaAndCHP}, it is clear that the distribution from scalar decay will solely depend on our effective coupling parameters $\CHP$ and $\CGamma$. The mass scales $M_1$ and $m_S$ only enter when calculating dimensionful quantities like the free-streaming horizon or the relic density. Furthermore, the discussion of Sec.~\ref{sec:NumericalAnalysisDW} shows that the suppression of the initial abundance, which \emph{does} depend on $M_1$, cf.~\equref{eq:SolutionExactCondensed}, is very weak -- such that the relic abundance of the combined production mechanism can be cast into the following form, to a very good approximation:
\begin{align}
 \sub{\Omega}{SD+DW}\left(M_1\right) \approx \DD{\sub{\Omega}{SD} \left(\CHP,\CGamma\right)}{M_1} M_1 + \sub{\Omega}{DW}\left[M_1,\left(\theta^e,\theta^\mu,\theta^\tau\right)\right].
 \label{eq:OmegaSDPlusDWApprox}
\end{align}
Here, $\DD{\sub{\Omega}{SD}}{M_1}$ is the relic abundance per sterile neutrino mass for the case of pure scalar decay (cf.~\cite[Fig.~3]{Merle:2015oja}).

If we choose, for instance, $\left(\theta^e,\theta^\mu,\theta^\tau\right) = \left(\sub{\theta}{max}\left(M_1\right),0,0\right)$ in \equref{eq:OmegaSDPlusDWApprox}, we can numerically solve for $M_1$ for a fixed set of $\left(\CHP,\CGamma \right)$, in such a way that $\sub{\Omega}{SD+DW}\left(M_1\right)$ matches the current best-fit values from Planck~\cite{Planck:2015xua}. These are the subcases \emph{$\beta$}, corresponding to scenarios where the correct abundance is achieved for \emph{the same} couplings $\left(\CHP,\CGamma \right)$ as in subcase $\alpha$, even when including the DW contribution, however paying the price of having to rescale the sterile neutrino mass.\footnote{Note that $\theta^e<\sub{\theta}{max}\left(M_1\right)$ in case 3 where a maximal mixing would already yield an overabundance. See caption of \tabref{tab:OverviewBenchmarkCases} for more details.} This case is useful whenever the parameter space is scanned in terms of the effective couplings, e.g., when aiming at producing new versions of Figs.~3, 8(a), or 8(b) in Ref.~\cite{Merle:2015oja}.

The value for the mass obtained via this procedure can then be mapped back to a case with $\left(\CHP',\CGamma\right)$, such that SD without DW yields the same mass when applying the relic abundance constraint (subcases \emph{$\gamma$}). Keeping $\CGamma$ constant guarantees that the production time of both scenarios is comparable, which is the most meaningful comparison when it comes to structure formation. The defining characteristics of subcases $\alpha$, $\beta$, and $\gamma$ are briefly summarised in \tabref{tab:cases}.

\begin{table}[t!]
 \centering
 \begin{tabular}{|c|c|c|c|c|}
 \hline 
  Subcase & Value of $\CHP$ & Value of $\CGamma$ & $\theta$ & Mass \\
  \hline \hline
  $\alpha$ & $\CHP$ & $\CGamma$ & $\theta = 0$ & matched to Planck data \\ 
  \hline 
  $\beta$ & $\CHP$ & $\CGamma$ & $\theta \neq 0$ & matched to Planck data \\ 
  \hline 
  $\gamma$ & $\CHP'$ & $\CGamma$ & $\theta = 0$ & same value as in $\beta$\\
    \hline 
 \end{tabular}
\caption{\label{tab:cases}Characteristics of the different subcases.}
\end{table}

\begin{figure}[htbp]
 \centering
    \includegraphics[width=\textwidth]{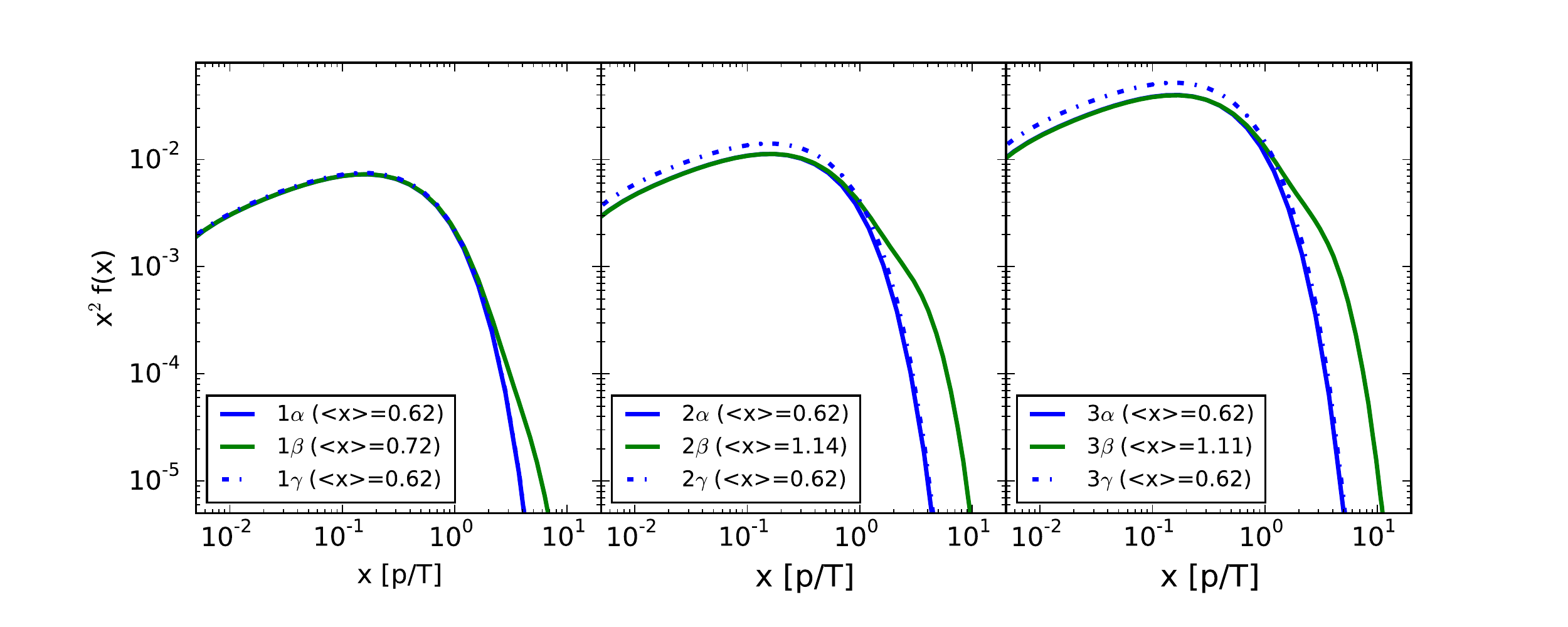}
  \caption{\label{fig:OverviewBenchmarkCases}Distribution functions and average momenta $\langle x \rangle$ corresponding to the benchmark-cases from \tabref{tab:OverviewBenchmarkCases}. \emph{From left to right}: Increasing values for $\CHP$ (corresponding to different sterile neutrino masses).}
\end{figure}

In \Figref{fig:OverviewBenchmarkCases}, we present the distribution functions for all (sub-)cases. In each panel, the solid blue curve corresponds to subcase~$\alpha$, while the dashed blue curve corresponds to subcase~$\beta$ and the green one to subcase~$\gamma$. We also quote the average momenta $\av{x}$ of the distribution. Note that the areas under the curves of subcase $\beta$ and under the ones for subcase $\gamma$ are identical within one and the same case by definition. 
%In what concerns the average momentum, we can see immediately that the changes are quite minor in most cases. 

\begin{table}
 \centering
 \begin{tabular}{|c|c|c|c|c|}
  \hline
  Case  & Description & $\CHP$ & $\CGamma$ & $M_1$ \\
  \hline \hline
  1$\alpha$ & SD only & $2.88\times10^{-2}$& $10^{3}$ & $\unit{7.1}{keV}$ \\
  \hline
  1$\beta$  & SD + maximal mixing&$2.88\times10^{-2}$& $10^{3}$ & $\unit{6.9}{keV}$ \\
  \hline
  1$\gamma$  & SD only &$2.96\times10^{-2}$& $10^{3}$ & $\unit{6.9}{keV}$ \\
  \hline \hline
  2$\alpha$  & SD only &$4.47\times10^{-2}$& $10^{3}$ & $\unit{4.5}{keV}$ \\
  \hline
  2$\beta$  & SD + maximal mixing&$4.47\times10^{-2}$& $10^{3}$ & $\unit{3.6}{keV}$ \\
  \hline
  2$\gamma$  & SD only&$5.59\times10^{-2}$& $10^{3}$ & $\unit{3.6}{keV}$ \\
  \hline \hline
  3$\alpha$  & SD only&$1.58\times10^{-1}$& $10^{3}$ & $\unit{1.3}{keV}$ \\
  \hline
  3$\beta$  & SD + mixing&$1.58\times10^{-1}$& $10^{3}$ & $\unit{1.0}{keV}$ \\
  \hline
  3$\gamma$  & SD only&$2.05\times10^{-1}$& $10^{3}$ & $\unit{1.0}{keV}$ \\
  \hline 
 \end{tabular}
  \caption{\label{tab:OverviewBenchmarkCases}Overview of the benchmark cases. All parameter sets are chosen such that the relic abundance of sterile neutrinos is in accordance with the best-fit value~\cite{Planck:2015xua}. Note that we do not assume \emph{maximal} mixing in agreement with X-ray constraints in case $3\beta$ since this would violate the overabundance bound. Still, the parameter $\CGamma'$ in case $3\gamma$ is chosen to reproduce the observed relic abundance with the same mass $M_1$ as in case $3\beta$.}
\end{table}

\begin{figure}
 \centering
 \includegraphics[width=\textwidth]{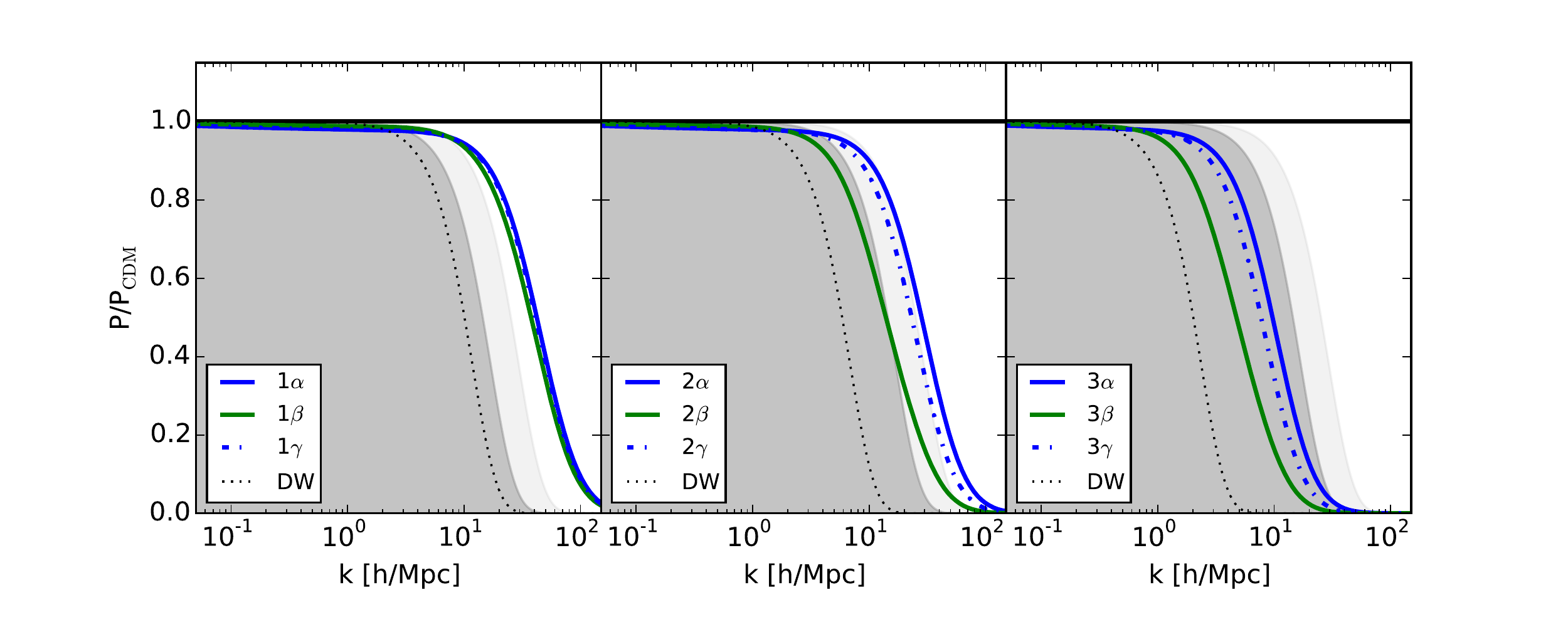}
  \caption{Relative power spectra of the benchmark-cases from \tabref{tab:OverviewBenchmarkCases}. The light- and dark-grey shaded areas illustrate the 2-$\sigma$ exclusion limits from Lyman-$\alpha$ observations~\cite{Viel:2013apy} and dwarf galaxy counts~\cite{Polisensky:2010rw}. The black dotted curves illustrate the pure DW case with the lower of the two particle masses given in \tabref{tab:OverviewBenchmarkCases}.}
  \label{fig:OverviewBenchmarkCases2}
\end{figure}

In order to assess the compatibility of the distribution functions shown in \Figref{fig:OverviewBenchmarkCases} beyond the simple approach using the average quantity of the free-streaming horizon, we have computed the linear power spectra $P$ for all 9 (sub-) cases and normalised them to the linear power spectra of a CDM analogue ($\sub{P}{CDM}$). The results are shown in \Figref{fig:OverviewBenchmarkCases2}, where we have adopted the colour coding of \Figref{fig:OverviewBenchmarkCases}. The figure clearly shows that, in all three benchmark cases, the spectrum from pure SD is always colder than the one with a DW component added. For masses around $\unit{7}{keV}$ (left panel) the difference is, however, very small, and all subcases are in agreement with the bounds from structure formation. For intermediate masses (central panel), the combined spectrum is only borderline compatible with structure formation, while the pure SD cases lie just a bit beyond the exclusion region. For smaller masses, around $\unit{1}{keV}$, the difference between the different cases is largest, as expected from the fact, that X-ray bounds are very weak in this regime. However, the corresponding subcases of case 3 (right panel) are clearly not compatible with Lyman-$\alpha$ observations.
This benchmark analysis shows that the additional DW component for a SD spectrum can usually be neglected, either because the effect is tiny to start with or because the model will be excluded with or without the additional DW component. Only in a very narrow range of parameters the additional DW component can possibly affect the validity of formerly borderline cases.\footnote{In all our benchmark cases, the combined spectrum is warmer than the pure SD spectrum. Note that the additional DW component can also \emph{cool} the spectrum, namely for small decay widths $\CGamma$, i.e.~late decays. However, in these cases even the DW-cooled spectrum is too warm to be compatible with structure formation, such that we did not consider it worthwhile to show such a case explicitly.} In the phenomenologically interesting region around $\unit{7}{keV}$, it is completely irrelevant whether the DW component is taken into account or not.

%%%%%%%%%%%%%%%%%%%%%%%%%%%%%%%%%%%%%%%%%%%%%%%%%%%%%%%%%%%%%%%%%%%%%%%%%%%%%%%%%%%%%%%%%%%%%%%%%%
\section{\label{sec:Conclusion}Conclusions}
%%%%%%%%%%%%%%%%%%%%%%%%%%%%%%%%%%%%%%%%%%%%%%%%%%%%%%%%%%%%%%%%%%%%%%%%%%%%%%%%%%%%%%%%%%%%%%%%%%

In this work, we have presented for the first time a fully comprehensive semi-analytical method for calculating the effect of the DW mechanism on \emph{any} initial distribution of sterile neutrinos. Treating the combination of DW and any other production mechanism in the most abstract way has allowed us to find a formalism where the interference effects between the initial spectrum and the DW effect can be put into an intuitive form.
The formalism yields an analytical expression of the combined momentum distribution function for every point in time (or temperature), the evaluation of which is however only feasible numerically.

Taking into account constraints from X-ray observations, we have used our method to show that the spectra of the preceding production mechanism and DW production can simply be added (after being correctly redshifted) to a very good approximation: interference effects are of the order of a few percent at most. We also used our approach to assess the quality of the common approximation of a suppressed thermal shape of the DW spectrum. We have seen that the estimates with suppressed thermal shape are quite sensitive to the definition of the average number of degrees of freedom during production. Two meaningful choices of this average clearly underestimate the abundance for a given mass and mixing angle. They also underestimate the weight of high momenta in the sterile neutrino spectrum, which implies that the limits previously obtained from structure formation considerations are in fact too weak.

Furthermore, we have set limits on the maximal fraction of sterile neutrino DM produced from the DW mechanism using our numerical results and astrophysical constraints. In a conservative analysis, DW can contribute about 50\% of the total DM abundance, however, only for a mass of about $\unit{2}{keV}$. This results sets tight constraints on earlier and possibly too optimistic interpretations of the data.

We finally applied our method to the example case of an initial sterile neutrino abundance produced from the decay of a scalar singlet particle. Since a pure scalar decay spectrum of sterile neutrinos is not perfectly cold, the additional DW contribution must in any case be smaller than the 50\% quoted above. In this scenario, we find that the maximal contributions of DW to the initial spectrum lie in a negligible parameter range, where the resulting combined spectrum would be excluded by structure formation anyway. In regions where the remaining combined spectrum does not violate structure formation bounds, the effect of DW is negligibly small, which justifies the assumption of neglecting the DW contribution for scalar decay production as made in~\cite{Merle:2015oja}. In the phenomenologically interesting region of a sterile neutrino of $M_1=\unit{7.1}{keV}$  we find that it is indeed \emph{irrelevant} whether DW is accounted for or not if the scalar decays fast enough not to violate bounds from structure formation. Comparing the linear transfer function of this case to current bounds from Lyman-$\alpha$ observations, we showed that scalar decay is indeed a viable production mechanism for  keV sterile neutrinos.

%%%%%%%%%%%%%%%%%%%%%%%%%%%%%%%%%%%%%%%%%%%%%%%%%%%%%%%%%%%%%%%%%%%%%%%%%%%%%%%%%%%%%%%%%%%%%%%%%%
\section*{Acknowledgements}
%%%%%%%%%%%%%%%%%%%%%%%%%%%%%%%%%%%%%%%%%%%%%%%%%%%%%%%%%%%%%%%%%%%%%%%%%%%%%%%%%%%%%%%%%%%%%%%%%%

AM would like to thank D.~B\"odeker for useful discussions. AM furthermore acknowledges partial support by the European Union FP7 ITN-INVISIBLES (Marie Curie Actions, PITN-GA-2011-289442). AS is supported by the Synergia project EUCLID from the Swiss National Science Foundation. MT would like to thank Mikko Laine for helpful comments. Furthermore, he acknowledges financial support from the Studienstiftung des deutschen Volkes and from the IMPRS-EPP.

%%%%%%%%%%%%%%%%%%%%%%%%%%%%%%%%%%%%%%%%%%%%%%%%%%%%%%%%%%%%%%%%%%%%%%%%%%%%%%%%%%%%%%%%%%%%%%%%%%
\appendix
%%%%%%%%%%%%%%%%%%%%%%%%%%%%%%%%%%%%%%%%%%%%%%%%%%%%%%%%%%%%%%%%%%%%%%%%%%%%%%%%%%%%%%%%%%%%%%%%%%

%%%%%%%%%%%%%%%%%%%%%%%%%%%%%%%%%%%%%%%%%%%%%%%%%%%%%%%%%%%%%%%%%%%%%%%%%%%%%%%%%%%%%%%%%%%%%%%%%%
\section{\label{app:A:FormalSolution}Appendix~A: Technical details on the formal solution of Eq.~(4)}
%%%%%%%%%%%%%%%%%%%%%%%%%%%%%%%%%%%%%%%%%%%%%%%%%%%%%%%%%%%%%%%%%%%%%%%%%%%%%%%%%%%%%%%%%%%%%%%%%%

\renewcommand{\theequation}{A-\arabic{equation}}
% redefine the command that creates the equation no.
\setcounter{equation}{0}  % reset counter 

%%%%%%%%%%%%%%%%%%%%%%%%%%%%%%%%%%%%%%%%%%%%%%%%%%%%%%%%%%%%%%%%%%%%%%%%%%%%%%%%%%%%%%%%%%%%%%%%%%
\subsection{\label{app:A:FormalSolution:DWChoiceNR} On the number of sterile neutrinos $\boldsymbol{n_R}$}
%%%%%%%%%%%%%%%%%%%%%%%%%%%%%%%%%%%%%%%%%%%%%%%%%%%%%%%%%%%%%%%%%%%%%%%%%%%%%%%%%%%%%%%%%%%%%%%%%%

In the general case with arbitrary $n_R$, \equref{eq:LagrangianTerms} takes the form
\begin{align}
 \Lag \supset -L_\alpha \tilde{H} y_{\alpha i} N_{i} - \frac{1}{2} \overline{\conj{N_i}} M_{ij} N_{j} + \hc \;,
 \label{eq:LagrangianTermsNR}
\end{align}
where the Majorana mass matrix $M_{ij}$ must be symmetric to fulfill the Majorana condition and can therefore always be diagonalised unitarily. There is no theoretical upper limit on $n_R$. If active-neutrino masses are to be explained by a seesaw mechanism, $n_R\geq2$ is needed to accommodate the two distinct mass scales $\sub{\Delta m^2}{atm}$ and $\sub{\Delta m^2}{sol}$ observed by neutrino oscillation experiments~\cite{Schechter:1980gr,Xing:2007uq}. 

Before discussing the physical aspects of DW, it is worthwhile to work out the number of parameters in a general $3+n_R$ framework: choosing a basis where $M_{ij}$ is diagonal, we are left with the $n_R$ real eigenvalues (if $CP$ is conserved) and $6n_R$ real parameters in the complex Yukawa matrix $y_{\alpha i}$, which can be subdivided into $3n_R$ mixing angles and equally many phases. Three of these phases can be absorbed into the charged lepton fields and hence we are left with $n_R$ Majorana masses, $3n_R$ mixing angles and $\left(3n_R -3\right)$ phases, i.e.\ $\left(7 n_R -3\right)$ free parameters in total. In this case, the mixing angles are given by
\begin{align}
 \theta^{\alpha i} = \frac{y_{\alpha i} \sub{v}{EW}}{M_i}\,,
 \label{eq:DefineMixingAnglesNR}
\end{align}
again assuming them to be sufficiently small.

%%%%%%%%%%%%%%%%%%%%%%%%%%%%%%%%%%%%%%%%%%%%%%%%%%%%%%%%%%%%%%%%%%%%%%%%%%%%%%%%%%%%%%%%%%%%%%%%%%
\subsection{\label{app:DW} The DW mechanism}
%%%%%%%%%%%%%%%%%%%%%%%%%%%%%%%%%%%%%%%%%%%%%%%%%%%%%%%%%%%%%%%%%%%%%%%%%%%%%%%%%%%%%%%%%%%%%%%%%%
 
To recall the Dodelson-Widrow (DW) mechanism (see~\cite{Dodelson:1993je} or also~\cite{Barbieri:1989ti,Kainulainen:1990ds} for earlier works) in a nutshell, it is basically some freeze-in~\cite{McDonald:2001vt,Hall:2009bx} type of DM-production. This means that the DM particles are interacting so feebly that they never enter thermal equilibrium. Instead, they start of with a certain zero or non-zero abundance and are gradually produced in the early Universe as long as they are kinematically accessible. In the case of sterile neutrinos, their small admixtures $\theta^\alpha$ with the active-neutrino sector cause them to be produced in the small fraction $|\theta^\alpha|^2$ of reactions where a vertex of flavour $\alpha$ happens to produce the keV-scale mass eigenstate $N_1$ instead of one of the three light mass eigenstates $\nu_{1,2,3}$.

The Boltzmann equation describing this type of production has been given in Eq.~\eqref{eq:BoltzmannDWAbstract}, and the part of the right-hand side which we abbreviated as $h(p,T)$ is explicitly given by\footnote{In the case of a primordial lepton number asymmetry there would be further potentials~\cite{Abazajian:2001nj}.}
\begin{equation}
 h(p,T) = \frac{1}{8} \frac{\Gamma_\alpha(p,T) \Delta^2(p) \sin^2(2\theta^\alpha)}{\Delta^2(p) \sin^2(2\theta^\alpha) + [\Gamma_\alpha(p,T)/2]^2(p) + [\Delta(p) \cos(2\theta^\alpha) - V_\alpha(p,T)]^2},
 \label{eq:h-explicit}
\end{equation}
where $\Gamma_\alpha(p,T)$ are the interaction rates of active (anti-) neutrinos of flavour $\alpha$, $V_\alpha(p,T)$ is the background potential for active (anti-) neutrinos of flavour $\alpha$, $\Delta(p) = (M_1^2 - m_\nu^2)/(2p) \simeq M_1^2/(2p)$ (where $m_\nu$ denotes any light neutrino mass which is always negligible compared to $M_1$), and $\theta^\alpha$ is the \emph{active-sterile mixing angle} for neutrinos of flavour $\alpha$, i.e., the fraction of the sterile neutrino mass eigenstate contained in the active flavour $\alpha$. The basic difficulty is to compute the interaction rates $\Gamma_\alpha(p,T)$ accurately, and to have a reliable expression for the potential $V_\alpha(p,T)$.

%%%%%%%%%%%%%%%%%%%%%%%%%%%%%%%%%%%%%%%%%%%%%%%%%%%%%%%%%%%%%%%%%%%%%%
\subsubsection{\label{sec:interactions}The interaction rates $\boldsymbol{\Gamma_\alpha(p,T)}$}
%%%%%%%%%%%%%%%%%%%%%%%%%%%%%%%%%%%%%%%%%%%%%%%%%%%%%%%%%%%%%%%%%%%%%%

\begin{figure}
\begin{center}
\includegraphics[width=8cm]{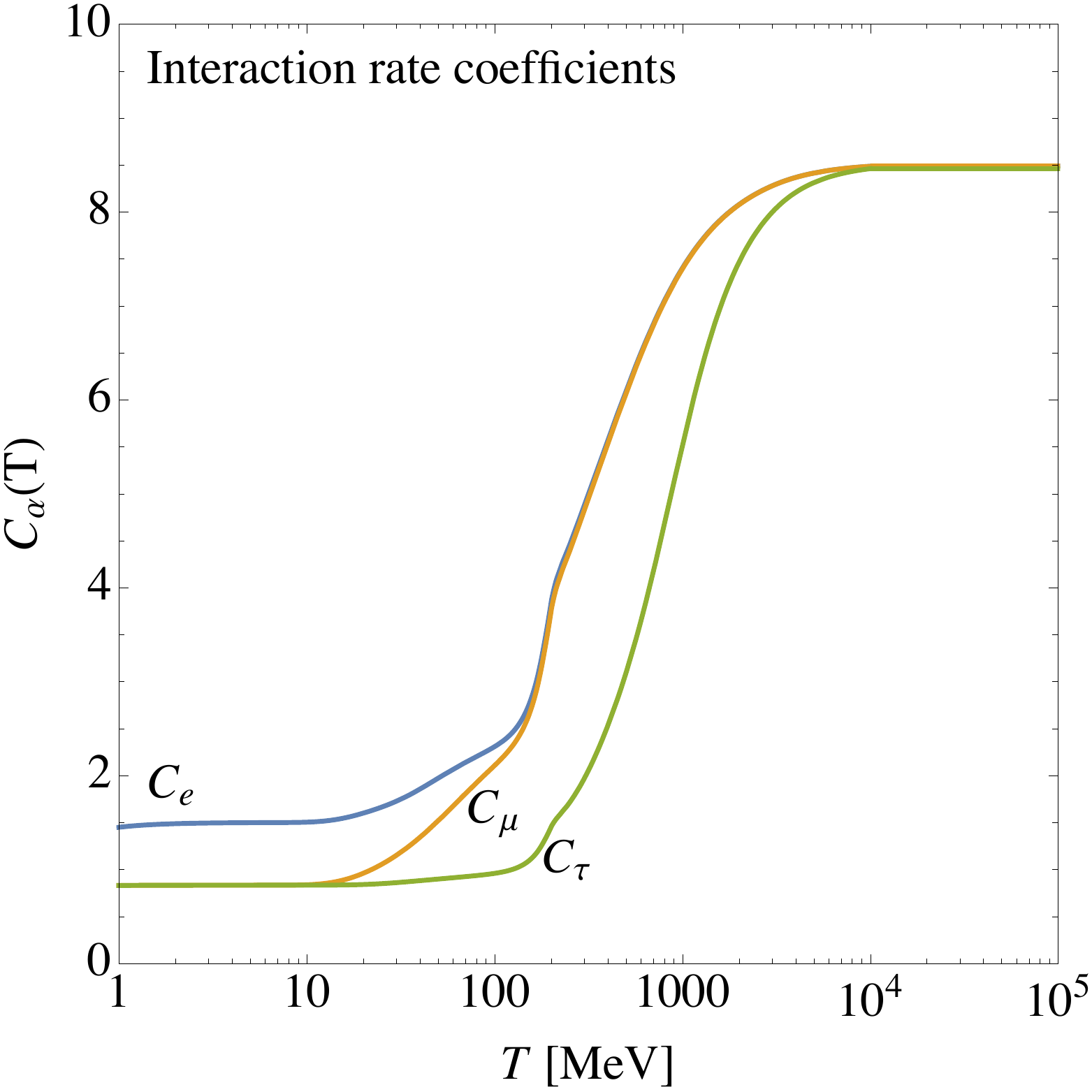}
\end{center}
\caption{\label{fig:C_alpha} The evolution of the coefficients $C_\alpha (T)$ with the temperature~\cite{Asaka:2006nq,Asaka:2006rw,Laine:2008pg}.}
\end{figure}

The basic form of the interaction rate $\Gamma_\alpha(p,T)$ is given by~\cite{Abazajian:2001nj,Kishimoto:2008ic,Hernandez:2014fha}:
\begin{equation}
 \Gamma_\alpha(p,T) = C_\alpha (T) G_F^2 p T^4,
 \label{eq:rate_1}
\end{equation}
where $G_F = 1.166 \cdot 10^{-5}~{\rm GeV}^{-2}$ is the Fermi constant, $p$ is the sterile neutrino momentum, and $C_\alpha (T)$ are temperature-dependent functions. These functions depend on the details of the dynamics in the plasma in the early Universe, and while first computations have been presented quite a while ago~\cite{Notzold:1987ik}, a detailed 2-loop calculation was performed only later on~\cite{Asaka:2006nq}. Most importantly, the results obtained in this reference have been recently made available in machine-readable numerical data files.\footnote{See \url{http://www.laine.itp.unibe.ch/neutrino-rate/}, where the interaction rates are given as twice the imaginary parts of the self-energies.} The values of $C_\alpha (T)$ as functions of the temperature are depicted in Fig.~\ref{fig:C_alpha}. Note that the functions as displayed here include the QCD contributions which are neglected in some references, like in Ref.~\cite{Hernandez:2014fha}. In particular this information complements the treatment presented in Ref.~\cite{Abazajian:2001nj}, where the interaction rates were only presented for a relatively narrow temperature range.

%%%%%%%%%%%%%%%%%%%%%%%%%%%%%%%%%%%%%%%%%%%%%%%%%%%%%%%%%%%%%%%%%%%%%%
\subsubsection{\label{sec:potentials}The potentials $\boldsymbol{V_\alpha(p,T)}$}
%%%%%%%%%%%%%%%%%%%%%%%%%%%%%%%%%%%%%%%%%%%%%%%%%%%%%%%%%%%%%%%%%%%%%%

Also the potentials $V_\alpha(p,T)$ are not often discussed in the generality required and, partially, the literature is plagued by unfortunate typos. We thus display the potentials, although known in principle, in full generality (see the discussions in~\cite{Abazajian:2001nj,Chu:2006ua}):
\begin{equation}
 V_\alpha(p,T) = \pm \sqrt{2} G_F \frac{2\zeta(3) T^3}{\pi^2} \frac{\eta_B}{4} - \frac{8\sqrt{2} G_F}{3 M_Z^2} \cdot 2 n_\alpha \langle E_\alpha \rangle - \frac{8\sqrt{2} G_F}{3 M_W^2} \cdot 2 n_{\alpha^\mp} \langle E_{\alpha^\mp} \rangle,
 \label{eq:potential_1}
\end{equation}
where the upper (lower) sign holds for neutrinos (anti-neutrinos), $\zeta(x)$ is the Riemann $\zeta$-function, and $\eta_B = 6.05\cdot 10^{-10}$ is the baryon asymmetry.\footnote{Note the discrepancy of a factor of 2 between Refs.~\cite{Abazajian:2001nj} and~\cite{Chu:2006ua}.} Here, the number densities and average energies for the neutrinos or anti-neutrinos of flavour $\alpha$ are given by
\begin{equation}
 n_\alpha = \frac{2\zeta(3) T^3}{4\pi^2}\ \ \ {\rm and}\ \ \ \langle E_\alpha \rangle = \frac{7\pi^4 T}{180 \zeta(3)},
 \label{eq:potential_2}
\end{equation}
where we have neglected the chemical potentials and we have set the active neutrino masses to zero. Their counterparts for the charged leptons are given by
\begin{equation}
 n_{\alpha^\mp} = \frac{T^3}{2 \pi^2} \cdot I_2 (m/T)\ \ \ {\rm and}\ \ \ \langle E_{\alpha^\mp} \rangle = T \frac{I_3 (m/T)}{I_2 (m/T)},
 \label{eq:potential_3}
\end{equation}
where the integrals
\begin{equation}
 I_n (x) \equiv \int\limits_0^\infty \frac{y^n}{e^{\sqrt{x^2+y^2}}-1} \diffd y
 \label{eq:potential_4}
\end{equation}
are evaluated numerically. Effectively, the contribution of the charged lepton of flavour $\alpha$ is zero for $T < m_\alpha$.

%%%%%%%%%%%%%%%%%%%%%%%%%%%%%%%%%%%%%%%%%%%%%%%%%%%%%%%%%%%%%%%%%%%%%%%%%%%%%%%%%%%%%%%%%%%%%%%%%%
\subsection{\label{app:A:FormalSolution:ConsitencyRelationProof} Proof of Eq.~(13)}
%%%%%%%%%%%%%%%%%%%%%%%%%%%%%%%%%%%%%%%%%%%%%%%%%%%%%%%%%%%%%%%%%%%%%%%%%%%%%%%%%%%%%%%%%%%%%%%%%%

In this appendix, we want to show all relevant steps to prove the consistency condition of the solution stated in \equref{eq:SolutionConservationRule}:
\begin{eqnarray}
  &&\mathrm{LHS}\equiv \mathcal{S}\left(\sub{T}{f},\sub{T}{ini},\sub{T}{f},p\right) \sub{f}{DW}\left(\sub{T}{f},\sub{T}{ini},p\right) \stackrel{!}{=} \nonumber\\
  &&\mathcal{S}\left(\sub{T}{f},T_3,\sub{T}{f},p\right) \Bigg[\mathcal{S}\left(\sub{T}{3},\sub{T}{ini},T_3,\frac{T_3}{\sub{T}{f}} \left(\frac{g_S\left(T_3\right)}{g_S\left(\sub{T}{f}\right)}\right)^{1/3}p\right) \sub{f}{DW}\left(T_3,\sub{T}{ini},\frac{T_3}{\sub{T}{f}} \left(\frac{g_S\left(T_3\right)}{g_S\left(\sub{T}{f}\right)}\right)^{1/3}p\right) + \nonumber\\
  &&\quad\quad\quad\quad\quad\quad\quad\quad\quad \sub{f}{DW}\left(\sub{T}{f},T_3,p\right)\Bigg] \equiv\mathrm{RHS}.
 \label{eq:SolutionConservationRuleLHSRHS}
\end{eqnarray}
To this end we introduce a few useful relations for the suppression factor $\mathcal{S}$, which follow directly from its definition, cf.~\equref{eq:Def:S}: 

\begin{equation}
 \mathcal{S}\left(T_a,T_b,T_c,p\right) =\mathcal{S}^{-1}\left(T_b,T_a,T_c,p\right) \quad\text{(Inversion),}
 \label{eq:SRelations1}
\end{equation}
\vspace{-0.1cm}
\begin{equation}
 \mathcal{S}\Bigg(T_b,T_a,\hat{T}, \frac{\hat{T}}{T_a} \Bigg(\frac{g_S\left(\hat{T}\right)}{g_S\left(T_c\right)}\Bigg)^{1/3}p\Bigg) = \mathcal{S}^{-1}\left(T_c,T_b,T_c,p\right) \mathcal{S}\left(T_c,T_a,T_c,p\right) \quad\quad \text{(general rescaling),}
 \label{eq:SRelations2} 
\end{equation}

\begin{equation}
 \mathcal{S}\Bigg(T_b,T_a,T_c, \frac{T_c}{T_a} \Bigg(\frac{g_S\left(T_c\right)}{g_S\left(T_a\right)}\Bigg)^{1/3}p\Bigg) = \mathcal{S}\left(T_b,T_a,T_a,p\right) \quad\quad \text{(particular rescaling),}
 \label{eq:SRelations3}
\end{equation}

\begin{equation}
 \mathcal{S}\left(T_a,T_b,T_d,p\right) \mathcal{S}\left(T_b,T_c,T_d,p\right) = \mathcal{S}\left(T_a,T_c,T_d,p\right) \quad\quad \text{(transitivity).}
 \label{eq:SRelations4}
\end{equation}

Let us start by manipulating the $\mathrm{RHS}$ of the equation:
{\footnotesize
\begin{align*}
 &\mathrm{RHS} =  \\
 &\stackrel{\eqref{eq:SRelations2}}{=}
 \mathcal{S}\left(\sub{T}{f},T_3,\sub{T}{f},p\right) 
 \left[
 \mathcal{S}^{-1}\left(\sub{T}{f},T_3,\sub{T}{f},p\right) 
 \mathcal{S}\left(\sub{T}{f},\sub{T}{ini},\sub{T}{f},p\right) 
 \sub{f}{DW}\left(T_3,\sub{T}{ini},\frac{T_3}{\sub{T}{f}}\left(\frac{g_S\left(T_3\right)}{g_S\left(\sub{T}{f}\right)}\right)^{1/3}\right)
 + 
 \sub{f}{DW}\left(\sub{T}{f},T_3,p\right)
 \right] 
%%%%%%%%%%%%%%%%%%%%%%%%%%%%%%%%%new step in proof %%%%%%%%%%%%%%%%%%%%%%%
 \\
 &\stackrel{\eqref{eq:Def:fDW}}{=} 
 \mathcal{S}\left(\sub{T}{f},\sub{T}{ini},\sub{T}{f},p\right)
 \int\limits_{T_3}^{\sub{T}{ini}}
 {
 \diffd T' \left(-1\right)
 \mathcal{S}^{-1}\left(T',\sub{T}{ini},T_3,\frac{T_3}{\sub{T}{f}}\left(\frac{g_S\left(T_3\right)}{g_S\left(\sub{T}{f}\right)}\right)^{1/3}p\right) 
 \left(h\sub{f}{th}\right)\left(T',\frac{T'}{T_3}\left(\frac{g_S\left(T'\right)}{g_S\left(T_3\right)}\right)^{1/3} \frac{T_3}{\sub{T}{f}}\left(\frac{g_S\left(T_3\right)}{g_S\left(\sub{T}{f}\right)}\right)^{1/3}p\right)
 } 
 \\
 & \quad\quad\quad + \mathcal{S}\left(\sub{T}{f},T_3,\sub{T}{f},p\right) 
 \int\limits_{\sub{T}{f}}^{T_3}
 {
 \diffd T' \left(-1\right)\mathcal{S}^{-1}\left(T',T_3,\sub{T}{f},p\right) 
 \left(h\sub{f}{th}\right)\left(T',\frac{T'}{\sub{T}{f}}\left(\frac{g_S\left(T'\right)}{g_S\left(\sub{T}{f}\right)}\right)^{1/3}p\right)
 } 
 \\
%%%%%%%%%%%%%%%%%%%%%%%%%%%%%%%%%new step in proof %%%%%%%%%%%%%%%%%%%%%%%
 &\stackrel{\eqref{eq:SRelations2}}{=} 
 \mathcal{S}\left(\sub{T}{f},\sub{T}{ini},\sub{T}{f},p\right) 
 \int\limits_{T_3}^{\sub{T}{ini}}
 {
 \diffd T' \left(-1\right) 
 \mathcal{S}^{-1}\left(\sub{T}{f},\sub{T}{ini},\sub{T}{f},p\right)  
 \mathcal{S}\left(\sub{T}{f},T',\sub{T}{f},p\right) 
 \left(h\sub{f}{th}\right)\left(T',\frac{T'}{\sub{T}{f}}\left(\frac{g_S\left(T'\right)}{g_S\left(\sub{T}{f}\right)}\right)^{1/3}p\right)
 } 
 \\
 & \quad\quad\quad + \mathcal{S}\left(\sub{T}{f},T_3,\sub{T}{f},p\right) 
 \int\limits_{\sub{T}{f}}^{T_3}
 {
 \diffd T' \left(-1\right) 
 \mathcal{S}^{-1}\left(T',T_3,\sub{T}{f},p\right) 
 \left(h\sub{f}{th}\right)\left(T',\frac{T'}{\sub{T}{f}}\left(\frac{g_S\left(T'\right)}{g_S\left(\sub{T}{f}\right)}\right)^{1/3}p\right)
 } 
 \\
%%%%%%%%%%%%%%%%%%%%%%%%%%%%%%%%%new step in proof %%%%%%%%%%%%%%%%%%%%%%%
 &\stackrel{\eqref{eq:SRelations4}}{=} 
 \int\limits_{T_3}^{\sub{T}{ini}}
 {
 \diffd T' \left(-1\right) 
 \mathcal{S}\left(\sub{T}{f},T',\sub{T}{f},p\right) 
 \left(h\sub{f}{th}\right)\left(T',\frac{T'}{\sub{T}{f}}\left(\frac{g_S\left(T'\right)}{g_S\left(\sub{T}{f}\right)}\right)^{1/3}p\right)
 } 
 \\
 & \quad\quad\quad + 
 \int\limits_{\sub{T}{f}}^{T_3}
 {
 \diffd T' \left(-1\right) 
 \mathcal{S}\left(\sub{T}{f},T',\sub{T}{f},p\right) 
 \left(h\sub{f}{th}\right)\left(T',\frac{T'}{\sub{T}{f}}\left(\frac{g_S\left(T'\right)}{g_S\left(\sub{T}{f}\right)}\right)^{1/3}p\right)
 } 
 \\
%%%%%%%%%%%%%%%%%%%%%%%%%%%%%%%%%new step in proof %%%%%%%%%%%%%%%%%%%%%%%
 &= \int\limits_{\sub{T}{f}}^{\sub{T}{ini}}{\diffd T' \left(-1\right) \mathcal{S}\left(\sub{T}{f},T',\sub{T}{f},p\right) \left(h\sub{f}{th}\right)\left(T',\frac{T'}{\sub{T}{f}}\left(\frac{g_S\left(T'\right)}{g_S\left(\sub{T}{f}\right)}\right)^{1/3}p\right)} \,.
 \\
%%%%%%%%%%%%%%%%%%%%%%%%%%%%%%%%%new step in proof %%%%%%%%%%%%%%%%%%%%%%%
\end{align*}}%

Let us now turn to the $\mathrm{LHS}$ of the equation, which just needs two simple steps:

{\footnotesize
\begin{align*}
 &\mathrm{LHS} 
 \stackrel{\eqref{eq:SRelations1}}{=} 
 \mathcal{S}\left(\sub{T}{f},\sub{T}{ini},\sub{T}{f},p\right)
 \int\limits_{\sub{T}{f}}^{\sub{T}{ini}}
 {
 \diffd T' \left(-1\right) \mathcal{S}\left(\sub{T}{ini},T',\sub{T}{f},p\right) 
 \left(h\sub{f}{th}\right)\left(T',\frac{T'}{\sub{T}{f}}\left(\frac{g_S\left(T'\right)}{g_S\left(\sub{T}{f}\right)}\right)^{1/3}p\right)
 } 
 \\
 &\stackrel{\eqref{eq:SRelations4}}{=} 
 \int\limits_{\sub{T}{f}}^{\sub{T}{ini}}
 {
 \diffd T' \left(-1\right) 
 \mathcal{S}\left(\sub{T}{f},T',\sub{T}{f},p\right) 
 \left(h\sub{f}{th}\right)\left(T',\frac{T'}{\sub{T}{f}}\left(\frac{g_S\left(T'\right)}{g_S\left(\sub{T}{f}\right)}\right)^{1/3}p\right)
 } 
 = \mathrm{RHS}\,. \quad \blacksquare	
\end{align*}
}%

%%%%%%%%%%%%%%%%%%%%%%%%%%%%%%%%%%%%%%%%%%%%%%%%%%%%%%%%%%%%%%%%%%%%%%%%%%%%%%%%%%%%%%%%%%%%%%%%%%
\subsection{\label{app:A:FormalSolution:ScalingBehaviour} Discussion of $\mathcal{S}$ and $\boldsymbol{\sub{f}{DW}}$}
%%%%%%%%%%%%%%%%%%%%%%%%%%%%%%%%%%%%%%%%%%%%%%%%%%%%%%%%%%%%%%%%%%%%%%%%%%%%%%%%%%%%%%%%%%%%%%%%%%

In general, the combination of Eqs.~\eqref{eq:Def:S} and~\eqref{eq:Def:fDW} with the definition of $h$ allow for the exact calculation of $\mathcal{S}$ and $\sub{f}{DW}$. Still, the computation can be numerically advanced and expensive, which is due to the complicated structure of the arguments in the functions $\sub{f}{th}$ and $h$, and also due to the rapidly varying shape of $h$ itself (cf.~\Figref{fig:h-variance}). Therefore, we want to state some simple (partially approximate) scaling relations of the solution which can prove helpful when calculating $\mathcal{S}$ and $\sub{f}{DW}$ for an extensive grid of mixing angles.

In the case of $\mathcal{S}$ this is simple. Since $h\propto\sin^2\theta$, one finds immediately:
\begin{align}
 \mathcal{S}\left(T_a,T_b,T_c,p,\Delta m^2,\theta\right) = \mathcal{S}\left(T_a,T_b,T_c,p,\Delta m^2,\sub{\theta}{ref}\right) ^{\sin^2\left(2\theta\right)/\sin^2\left(2\sub{\theta}{ref}\right)} \;,
 \label{eq:ScalingSOneAngle}
\end{align}
with $\sub{\theta}{ref}$ being some arbitrary (small) reference angle.
Taking into account three non-zero mixing angles, we find
\begin{align}
 \mathcal{S}\left(T_a,T_b,T_c,p,\Delta m^2,\left(\theta^e,\theta^\mu,\theta^\tau\right)\right) =
 \mathcal{S}\left(...,\sub{\theta^e}{ref}\right) ^{\frac{\sin^2\left(2\theta^e\right)}{\sin^2\left(2\sub{\theta^e}{ref}\right)}}
 \mathcal{S}\left(...,\sub{\theta^\mu}{ref}\right) ^{\frac{\sin^2\left(2\theta^\mu\right)}{\sin^2\left(2\sub{\theta^\mu}{ref}\right)}}
 \mathcal{S}\left(...,\sub{\theta^\tau}{ref}\right) ^{\frac{\sin^2\left(2\theta^\tau\right)}{\sin^2\left(2\sub{\theta^\tau}{ref}\right)}}
 \label{eq:ScalingSThreeAngles}
\end{align}
In the case of $\sub{f}{DW}$, we can only find an approximate relation:
\begin{align}
 \sub{f}{DW}\left(T_a,T_b,p,\Delta m^2,\left(\theta^e,\theta^\mu,\theta^\tau\right)\right) \approx \sum\limits_{i=e,\mu\tau}{\frac{\sin^2\left(\theta^i\right)}{\sin^2\left(\sub{\theta^i}{ref}\right)}} \sub{f}{DW}\left(T_a,T_b,T_c,\Delta m^2,\sub{\theta^i}{ref}\right) 
 \label{eq:ScalingfDW}
 \;.
\end{align}
This approximation is valid as long as the term converting sterile neutrinos into active ones is small, which is certainly the case for pure (i.e.~non-resonant) DW production without initial abundance. In this case, we get the simple scaling behaviour expected for the solution, being directly proportional to the sine-square of the active-sterile mixing angle.

%=============================================================================
\bibliographystyle{./apsrev}
\bibliography{MaxTotzauerBib}
%=============================================================================

\end{document}